\documentclass[a4paper,11pt]{article}
\pdfoutput=1 

\usepackage{jcappub} 

\usepackage{subcaption}
\usepackage{soul}
\usepackage{xcolor}
\sethlcolor{yellow} 

\usepackage[font={small}]{caption}
\usepackage[utf8]{inputenc}
\title{\boldmath Relativistic and non-Gaussianity contributions to the one-loop power spectrum}

\author[a,1]{Rebeca Martinez-Carrillo,\note{Corresponding author.}}
\author[b,c]{Josue De-Santiago,}
\author[d]{Juan Carlos Hidalgo,}
\author[a]{Karim A. Malik}

\affiliation[a]{Astronomy Unit, School of Physics and Astronomy, Queen Mary University of London,\\ Mile End Road, London, E1 4NS, United Kingdom.}
\affiliation[b]{Departamento de F\'isica, Centro de Investigaci\'on y de Estudios Avanzados del IPN, \\ A.P. 14-740, 07000 Ciudad de M\'exico, M\'exico.}
\affiliation[c]{C\'atedra - Consejo Nacional de Ciencia y Tecnolog\'ia, \\ Av. Insurgentes Sur 1582, 03940 Mexico City, M\'exico.}
\affiliation[d]{Instituto de Ciencias F\'isicas, Universidad Nacional Aut\'onoma de M\'exico,\\  62210, Cuernavaca, Morelos, M\'exico.}

\emailAdd{r.martinezcarrillo@qmul.ac.uk}
\emailAdd{jsantiago@fis.cinvestav.mx}
\emailAdd{hidalgo@icf.unam.mx}
\emailAdd{k.malik@qmul.ac.uk}

\abstract{We compute the one-loop density power spectrum including Newtonian and relativistic contributions, as well as the primordial non-Gaussianity contributions from $f_{\rm NL}$ and $g_{\rm NL}$ in the local configuration. To this end we take solutions to the Einstein equations in the long-wavelength approximation and provide expressions for the matter density perturbation at second and third order. These solutions have shown to be complementary to the usual Newtonian cosmological perturbations.  We confirm a sub-dominant effect from pure relativistic terms, manifested at scales dominated by cosmic variance, but find that a sizable effect of order one comes from $g_{\rm NL}$ values allowed by Planck-2018 constraints, manifested at scales probed by forthcoming galaxy surveys like DESI and Euclid. As a complement, we present the matter bispectrum at the tree-level including the mentioned contributions.}


\begin{document}
\maketitle
\flushbottom
\newpage
\section{Introduction}

The upcoming Stage IV experiments like DESI \cite{Aghamousa:2016zmz} and Euclid \cite{Laureijs:2011gra} aim to map the large-scale structure (LSS) of the Universe at high-precision. For many years the Newtonian standard perturbation theory (SPT) \cite{Makino:1991rp,Jain:1993jh,Bernardeau:2001qr} was sufficient to study the evolution of the matter content of the Universe. However, the Newtonian description of the Universe is only adequate for scales well inside the horizon and non-relativistic matter. The scales explored by the future surveys make a relativistic description of the dynamics of the Universe essential.
The analysis of the new data gathered by new experiments requires the use of sophisticated statistical tools. The power spectrum and bispectrum density fluctuations in the matter are key quantities to study the formation and composition of the LSS.

Recent analytic and numerical work began to include relativistic effects into the density power spectrum and bispectrum. In particular: relativistic corrections to the power spectrum \cite{Bonvin:2011bg,Challinor:2011bk}, one-loop relativistic corrections for the power spectrum and bispectrum on intermediate
scales using the weak field approximation
\cite{Castiblanco:2018qsd}, 
galactic non-linear bias from gravitational non-linearity
\cite{Bruni:2011ta, Bertacca:2015mca, Umeh:2019qyd, Umeh:2019jqg,  Calles:2019prs},
and N-body simulations using General Relativity (GR) 
\cite{Adamek:2016zes, Barrera-Hinojosa:2019mzo, Bentivegna:2015flc}. 

In order to account for relativistic contributions to observables, a complementary strategy is to integrate the signals produced by galaxies along the line of sight to determine the distance-redshift relation. The resulting distortions of redshift-space due to structure have been reported in terms of the number counts of galaxy clustering at linear order \cite{jeong1,durrer1,antony,chen,zalda}, and at second order \cite{cc1,cc2,durrer2,yoo1,yoo2,yoo3,nielsen,yoo5,Fuentes:2019nel}. Moreover, the observed bispectrum receives contributions from this relation as shown recently in  \cite{durrer2,Umeh:2016nuh,Jolicoeur:2017nyt,Clarkson:2018dwn,Mitsou:2019ocs,Maartens:2019yhx,Bertacca:2017dzm}.

One of the quantities that the community aims to constrain with the data released by the forthcoming surveys, is the primordial non-Gaussianity (PNG) generated during inflation. The inflationary epoch is crucial
for early structure formation and the later evolution of the LSS \cite{Bartolo:2005xa,dePutter:2015vga,Bartolo:2015qva}, which makes constraining the PNG such an exciting prospect (see e.g. Ref.~\cite{Desjacques:2010jw} for a review), as this offers the possibility to probe different inflationary scenarios (see also Refs.~\cite{Bartolo:2004if,Giannantonio:2011ya} for prospects of detection).
The effects of non-Gaussianity on the power spectrum have been previously studied in Refs.~\cite{Taruya:2008pg,Desjacques:2008vf}, although not from a relativistic point of view. In Ref.~\cite{Bruni:2011ta} the authors include relativistic corrections to the galaxy power spectrum including primordial non-Gaussianities from $f_{\rm NL}$ in the local configuration. 
In Ref.~\cite{Tellarini:2016sgp} the galaxy bispectrum in redshift space, including PNG, is calculated. In Refs.~\cite{DiDio:2016gpd,Koyama:2018ttg} the effective non-Gaussianity from relativistic corrections to the bispectrum of galaxies is determined.
 
The goal of this paper is to include contributions from the scalar sector of the full relativistic theory at second and third order,
as well as the primordial non-Gaussianity at the same orders (which can be easily included as an additional term in the chosen gauge; the synchronous-comoving gauge), and analyse their effects on the power spectrum at one-loop and the  tree-level bispectrum.

The paper is organised as follows: in section \ref{sec:Evolutioneqs} we review work previously done and present the evolution equations for the density contrast in synchronous-comoving gauge. We present its solutions up to third order using the gradient expansion. These solutions assume an Einstein-de Sitter Universe and are necessary for the computation of the one-loop power spectrum. In section \ref{sec:Fouriersol}, we present the Newtonian and relativistic solutions for the density contrast in Fourier space. Section \ref{sec:One-loop} is dedicated to the one-loop power spectrum, which is our main result. We provide complete analytical expressions for the one-loop power spectrum, along with numerical integrations, including the contributions to the one-loop power spectrum for the allowed values of $f_{\rm NL}$ and $g_{\rm NL}$ reported by Planck \cite{Akrami:2019izv}. For completeness, in section
\ref{sec:Treebis} we present the tree-level bispectrum, along with numerical solutions. Finally, in section \ref{sec:conclusions} we discuss our results in light of the forthcoming galaxy surveys. 

Throughout this paper we use the conformal time $\eta$, and denote derivatives with respect to $\eta$ with a prime. Greek indices $\mu$, $\nu$, range from 0 to 3, lower case Latin indices, $i$, $j$, and $k$, have the range 1, 2, 3.

\section{Evolution equations and relativistic density contrast solutions}
\label{sec:Evolutioneqs}

In this section we present the evolution equations for the density contrast in synchronous-comoving gauge, based on work previously done in Refs.~\cite{Gressel:2017htk,Bruni:2014xma,Bruni:2013qta}. The choice of this gauge provides a Lagrangian frame in GR, which is also suitable for defining local Lagrangian galaxy bias up to second order \cite{Bertacca:2015mca}. Our starting point is the general line element,
\begin{equation}
ds^2=a^2(\eta)[-(1+2\phi)d\eta^2+2\omega_{,i}d\eta dx^i+\gamma_{ij}dx^idx^j],
\end{equation}

\noindent where $a$ is the scale factor, $\eta$ is the conformal time, $\phi$ and $\omega$ are scalar metric perturbations and $\gamma_{ij}$ is the spatial metric. As we will work in the synchronous comoving gauge, we set $\phi=\omega_{,i}=0$.

As the matter content we consider an irrotational, pressureless fluid. Observers are comoving with the fluid, and as a consequence the four-velocity in the synchronous comoving gauge is $u_\mu=(-a,0,0,0)$. 

For the following fluid description, we define the deformation tensor,
\begin{equation}
{\vartheta}_\nu^\mu\equiv a{u^\mu}_{;\nu}-\mathcal{H}\delta_\nu^\mu,
\label{lineelement}
\end{equation}

\noindent where $\mathcal{H}=a'/a$ is the conformal Hubble scalar, the semicolon denotes covariant derivative  and the isotropic background expansion was removed. In the chosen gauge, the deformation tensor has only spatial components and is proportional to the extrinsic curvature $K_j^i$ of the conformal spatial metric $\gamma_{ij}$,

\begin{equation}
\vartheta_j^i=-K_j^i,
\end{equation}

\noindent where $K_j^i$ is given by
\begin{equation}
K_j^i\equiv-\frac{1}{2}\gamma^{ik}\gamma_{kj}'.
\end{equation}

The density field $\rho$ is defined as 
\begin{equation}
\rho(\mathbf x,\eta)=\bar\rho(\eta)+\delta\rho(\mathbf x,\eta)=\bar\rho(\eta)(1+\delta(\mathbf x,\eta)),
\end{equation}

\noindent where $\bar\rho(\eta)$ is the density in the background, $\delta\rho(\mathbf x,\eta)$ is a small perturbation and $\delta(\mathbf x,\eta)$ is the density contrast. 
The evolution of the density contrast $\delta(\mathbf x,\eta)$ is given by the continuity equation 
\begin{equation}
\delta'+(1+\delta)\vartheta=0,
\label{continuity}
\end{equation}

\noindent where $\vartheta=\vartheta_\alpha^\alpha$ is the trace of $\vartheta_\nu^{\mu}$.

The evolution for $\vartheta$ is given by the Raychaudhuri equation (more details of the derivation can be found in Refs.~\cite{Bruni:2013qta,Meures})
\begin{equation}
\vartheta'+\mathcal{H}\vartheta+\vartheta_j^i\vartheta_i^j+4\pi G a^2 \bar\rho\delta=0.
\label{Ray}
\end{equation}

The energy constraint is given by 

\begin{equation}
\vartheta^2-\vartheta_j^i\vartheta_i^j+4\mathcal{H}\vartheta+{}^{3}R=16\pi Ga^2\bar\rho\delta,
\label{energycons}
\end{equation}

\noindent where ${}^{3}R$ is the spatial Ricci scalar of the spatial metric $\gamma_{ij}$. In the following subsections we use two approaches to find solutions to the evolution equations. 

\subsection{Cosmological perturbation theory}
\label{perturbation}
In order to show how cosmological perturbation theory is used to find the evolution of the density contrast, we present in this section the solutions to first order. The line element (\ref{lineelement}) is equivalent to a spatially flat FLRW background with a perturbed spatial metric (in synchronous-comoving gauge), and hence we can expand $\gamma_{ij}$ as in terms of the scalar metric potentials $\psi$ and $\chi$ as
\begin{align}
\label{pert_metric}  
\nonumber
\gamma_{ij}&=\delta_{ij}+\gamma_{ij}^{(1)}+\frac{1}{2}\gamma_{ij}^{(2)}+...\\
 &=(1-2\psi^{(1)}-\psi^{(2)})\delta_{ij}+\chi_{ij}^{(1)}+\frac{1}{2}\chi_{ij}^{(2)}+...
\end{align}

\noindent where 
\begin{equation}
\chi_{ij}=\left(\partial_i\partial_j-\frac{1}{3}\delta_{ij}\nabla^2\right)\chi.
\end{equation}

The density contrast is decomposed as
\begin{equation}
\delta=\delta^{(1)}+\frac{1}{2}\delta^{(2)}+\frac{1}{6}\delta^{(3)}+...
\end{equation}

For the case of the first order solutions for the density contrast, we combine the first order of the continuity equation (\ref{continuity}) and the Raychaudhuri equation (\ref{Ray}) at first order,  to obtain the first order density contrast evolution equation
\begin{equation}
\delta^{(1)''}+\mathcal{H}\delta^{(1)'}-\frac{3}{2}\mathcal{H}^2\Omega_m\delta^{(1)}=0.
\end{equation}

 From the first order energy constraint equation (\ref{energycons}), combined with the first order continuity equation (\ref{continuity}) we obtain
 \begin{equation}
 4\mathcal{H}\delta^{(1)'}+6\mathcal{H}^2\Omega_m\delta^{(1)}-{}^{3}R^{(1)}=0,
 \label{deltaprime}
 \end{equation}
 
\noindent combining the time derivative of the  Eq.(\ref{deltaprime}) and using the first order of Eqs. (\ref{continuity}) and (\ref{Ray}) we find an equation for $R^{(1)}$ given by
 \begin{equation}
 {}^{3}R^{(1)'}=0.
 \end{equation}
 
 The general solution for a second order differential equation, will be composed of a linear combination of a growing mode  and a decaying mode 
 \begin{equation}
 \delta^{(1)}(\eta,\mathbf x)=C_{+}(\mathbf x)D_{+}(\eta)+C_{-}(\mathbf x)D_{-}(\eta).
 \end{equation}
 
 Since we choose to work in an Einstein-de Sitter universe, the decaying mode solution is negligible and from now on we take a solution of the form
 \begin{equation}
\delta^{(1)}(\eta,\mathbf x)=C(\mathbf x)D_{+}(\eta),
\label{soldelta}
 \end{equation}

\noindent where $C(\mathbf x)$ will be given by \cite{Bruni:2013qta}
 \begin{equation}
C(\mathbf x)=\frac{{}^{3}R^{(1)}}{10\mathcal{H}_{IN}^2D_{+IN}},
\end{equation}
 
 \noindent $D_{+}$ is the growth factor, and the subscript ``$IN$" denotes a time early in the matter dominated era.

At first order in perturbative, for an unspecified gauge, the spatial Ricci scalar, is 

\begin{equation}
{}^{3}R^{(1)}=4\nabla^2\left(\psi^{(1)}+\frac{1}{6}\nabla^2\chi^{(1)}\right).
\label{Riccist}
\end{equation}
\noindent From Eq. (\ref{Riccist}) in the comoving gauge, we can identify the comoving curvature perturbation, $\mathcal{R}_c$, \cite{Malik:2008im,Carrilho:2015cma}
\begin{equation}
\mathcal{R}_c=\psi^{(1)}_c+\frac{1}{6}\nabla^2\chi^{(1)}_c.
\end{equation}

The comoving curvature perturbation is related with the curvature perturbation on the  uniform-density gauge as (see for example Ref. \cite{Malik:2008im})
\begin{equation}
\zeta^{(1)}\equiv -\psi^{(1)}-\frac{1}{6}\nabla^2\chi^{(1)}-\frac{\mathcal{H}}{\rho'}\delta\rho^{(1)}=-\mathcal{R}_c+\frac{1}{3}\delta^{(1)},
\end{equation}

\noindent and at early times and large scales $\zeta^{(1)}$ and $\mathcal{R}_c$ are approximately equal:
\begin{equation}
\zeta^{(1)}\simeq-\mathcal{R}_c.
\label{Rczeta}
\end{equation}
Substituting Eq.~(\ref{Rczeta}) into Eq.~(\ref{Riccist}), we write the first order solution for  the density contrast as 
\begin{equation}
\delta^{(1)}=\frac{D_{+}(\eta)}{10\mathcal{H}^2_{IN}D_{+IN}}\left(-4\nabla^2\zeta^{(1)}\right),
\label{delta1N}
\end{equation}

\noindent where the growth factor in Einstein-de Sitter is\footnote{The order by order correspondence between the density contrast and the curvature perturbation means that $\delta^{(1)}$ represents a Gaussian field.}
\begin{equation}
D_{+}=\frac{D_{+IN}\mathcal{H}_{IN}^2}{\mathcal{H}^2},
\end{equation}

\noindent with  $D_{+IN}=1$ and $\mathcal{H}_{IN}=\mathcal{H}_0$, where $\mathcal{H}_0$ is the conformal Hubble parameter at present time.  These choices are made to recover the standard Newtonian solutions. 

\subsection{Gradient expansion approach}

In section \ref{perturbation} we presented the first order equations and solutions for the density contrast using cosmological perturbation theory, in this section we present the solutions for the second and third order equations using a different approach, the gradient expansion, that leads to the same equations and solutions obtained using the perturbative treatment. Instead of using the expansion Eq.~(\ref{pert_metric}), we can also write the spatial metric as \cite{Salopek:1990jq,Lyth:2004gb}
\begin{equation}
g_{ij}=a^2\gamma_{ij}=a^2e^{2\zeta}\check\gamma_{ij},
\end{equation}
\noindent where $\zeta$ is the curvature perturbation on uniform density hypersurfaces. 

The initial conditions for perturbations are set in the inflationary epoch. After this period, the curvature perturbation $\zeta$ is almost scale-invariant and remains constant (see for example Ref. \cite{Lyth:2009zz}). As a consequence is it possible to consider small initial inhomogeneities on large scales, allowing for a gradient expansion \cite{Lifshitz:1963ps, Tomita:1975kj, Rampf:2012pu, Rampf:2013ewa, Deruelle:1994iz}. In this long-wavelength approximation the spatial gradients are small compared to time derivatives. Using this approximation we find  
\begin{equation}
\delta\sim \vartheta\sim {}^{3}R\sim \nabla^2,
\end{equation}

\noindent and using this approximation with the continuity (\ref{continuity}) and energy constraint equations (\ref{energycons}), lead us back to the Eq.~(\ref{deltaprime}).

On large scales, and only considering scalars,  the conformal metric can be approximated as $\check\gamma_{ij}\simeq \delta_{ij}$. As a consequence of this simplified spatial metric, the Ricci scalar $R$ is a nonlinear function of the curvature perturbation $\zeta$ only, taking the form \cite{{Gressel:2017htk},{Bruni:2014xma},{Wald:1984rg}}
\begin{equation}
{}^{3}R=-4\nabla^2\zeta+\sum_{m=0}^{\infty}\frac{\left(-2\right)^{m+1}}{(m+1)!}\left[(m+1)(\nabla\zeta)^2-4\zeta\nabla^2\zeta\right]\zeta^m.
\end{equation}

This expansion for $R$ will allow us to obtain solutions for the density contrast to higher orders. In this paper we are interested in solutions up to third order. The third order corrections are obtained after expanding $R$ up to $m=1$ and are given by
\begin{equation}
{}^{3}R=-4\nabla^2\zeta+(-2)[(\nabla\zeta)^2-4\zeta\nabla^2\zeta]+2[2(\nabla\zeta)^2-4\zeta\nabla^2\zeta]\zeta.
\label{Rm1}
\end{equation}

 The curvature perturbation can be expanded in terms of a Gaussian random field $\zeta^{(1)}$ as
\begin{equation}
\zeta=\zeta^{(1)}+\frac{3}{5}f_{\rm NL}\zeta^{(1)2}+\frac{9}{25}g_{\rm NL}\zeta^{(1)3},
\label{zetaexpansion}
\end{equation}

\noindent where $f_{\rm NL}$ and $g_{\rm NL}$ are the non-Gaussian parameters at first and second order respectively.
 After substituting Eq.~(\ref{zetaexpansion}) into Eq.~(\ref{Rm1}), we get an expression for the Ricci scalar, that will allow us to find the density contrast solutions
\begin{align}
 \label{Riccifnlexpansion}   
\nonumber
{}^{3}R\simeq-4\nabla^2\zeta^{(1)}&+\left(\nabla\zeta^{(1)}\right)^2\left[-2-\frac{24}{5}f_{\rm NL}\right]+\zeta^{(1)}\nabla^2\zeta\left[-\frac{24}{5}f_{\rm NL}+8\right]\\
              &+\zeta^{(1)}\left(\nabla\zeta^{(1)}\right)^2\left[-\frac{216}{25}g_{\rm NL}+\frac{24}{5}f_{\rm NL}+4\right]\\
              &+\zeta^{(1)2}\nabla^2\zeta^{(1)}\left[-\frac{108}{25}g_{\rm NL}+\frac{72}{5}f_{\rm NL}-8\right]+\mathcal{O}(\zeta^{(1)4}).
              \nonumber
\end{align}

From Eq.~(\ref{Riccifnlexpansion}) it is straightforward to see that solutions to first order in the gradient expansion agree with the ones produced using the perturbation theory treatment. 

In a similar way to the first order, using the continuity equation (\ref{continuity}), along  with the energy constraint equation (\ref{energycons}), the  second order  evolution equation of $\delta$ will be given by
\begin{equation}
4\mathcal{H}\delta^{(2)'}+6\mathcal{H}^2\Omega_m\delta^{(2)}-{}^{3}R^{(2)}=2\vartheta^{(1)2}-2\vartheta^{(1)i}_j\vartheta_i^{(1)j}-8\mathcal{H}\delta^{(1)}\vartheta^{(1)},
\end{equation}

\noindent using
\begin{equation}
{}^{3}R^{(2)'}=-4\vartheta^{(1)i}_jR_i^{(1)j}.
\end{equation}

As shown in the Ref. \cite{Bruni:2013qta}, the solution for these equations is composed of an homogeneous and a particular solution (labelled with subscripts ``$h$" and ``$p$" respectively) of the form
\begin{equation}
\label{partplushom}
\delta^{(2)}=\delta_h^{(2)}+\delta_p^{(2)}, \quad {}^{3}R^{(2)}={}^{3}R_h^{(2)}+{}^{3}R_p^{(2)}, 
\end{equation}

\noindent where the particular solution recovers the Newtonian density contrast obtained within the Newtonian standard perturbation theory formalism and the homogeneous solution corresponds to the relativistic contributions to the density contrast also presented in Ref.\cite{Bruni:2013qta}\footnote{Expressions for the relativistic contributions in the Lagrangian perturbation formalism have also been reported in \cite{Christopherson:2015ank}.}.

Thus, using the expansion for the Ricci scalar given in Eq.~(\ref{Riccifnlexpansion}) up to second order ($m=0$), the homogeneous solution for the  second order of the density contrast is 
\begin{equation}
\frac{1}{2}\delta^{(2)}=\frac{D_{+}(\eta)}{10\mathcal{H}^2_{IN}D_{+IN}}\frac{24}{5}\Bigg[-(\nabla\zeta^{(1)})^2\bigg(\frac{5}{12}+f_{\rm NL}\bigg)+\zeta^{(2)}\nabla^2\zeta^{(1)}\bigg(\frac{5}{3}-f_{\rm NL}\bigg)\Bigg],
\label{Reldelta2}
\end{equation}

\noindent in analogous way the homogeneous  third order solution for the density contrast is 
\begin{align}
\label{Reldelta3}
\nonumber
\frac{1}{6}\delta^{(3)}=\frac{D_{+}(\eta)}{10\mathcal{H}^2_{IN}D_{+IN}}\frac{108}{25}\Bigg[&2\zeta^{(1)}(\nabla\zeta^{(1)})^2\bigg(-g_{\rm NL}+\frac{5}{9}f_{\rm NL}+\frac{25}{54}\bigg)\\
+&\zeta^{(1)2}\nabla^2\zeta^{(1)}\bigg(-g_{\rm NL}+\frac{10}{3}f_{\rm NL}-\frac{50}{27}\bigg)\Bigg],
\end{align}

\noindent which slightly differs from the 
expression provided in Ref.~\cite{Gressel:2017htk}. We are interested in the new effects to the one-loop power spectrum due to Newtonian and relativistic contributions focusing on the derivation of the relativistic solutions for the density contrast, since the Newtonian solutions are well known (see e.g. \cite{Makino:1991rp,Jain:1993jh,Bernardeau:2001qr,Carlson:2009it}).

\section{Complete density contrast solutions in Fourier space}
\label{sec:Fouriersol}
In this section we present the complete solutions for the density contrast in Fourier space, these solutions consider both Newtonian and relativistic contributions.\footnote{In this paper we follow this Fourier convention $A(\mathbf x)=\int \frac{d^3\mathbf k}{(2\pi)^3} \widetilde {A}(\mathbf k)e^{i\mathbf k\cdot\mathbf x}$, and $\widetilde{A}(\mathbf k)=\int d^3\mathbf x A(\mathbf x)e^{-i\mathbf k\cdot\mathbf x}$.}

In Fourier space the second order density contrast is defined by 

\begin{equation}
\frac{\delta^{(2)}(\mathbf k,\eta)}{2}=\int\frac{d^3\mathbf k_1d^3\mathbf k_2}{(2\pi)^3}\delta_D(\mathbf k-\mathbf k_1-\mathbf k_2)\mathcal{F}^{(2)}(\mathbf k_1,\mathbf k_2,\eta)\delta^{(1)}(\mathbf k_1,\eta)\delta^{(1)}({\mathbf k_2},\eta),
\label{delta2N}
\end{equation}

\noindent the kernel $\mathcal{F}^{(2)}(\mathbf k_1,\mathbf k_2,\eta)$ is given by \footnote{The kernels presented in this section are symmetrized, this results from the sum of $\mathcal{F}^{(n)}$ with all possible permutations of $\mathbf{k}_i$, the symmetrized kernels are written in calligraphic font.}
\begin{equation}
\mathcal{F}^{(2)}(\mathbf k_1,\mathbf k_2,\eta)=\mathcal{F}_{N}^{(2)}(\mathbf k_1,\mathbf k_2,\eta)+\mathcal{F}_{R}^{(2)}(\mathbf k_1,\mathbf k_2,\eta),
\label{Kernels2}
\end{equation}

\noindent where  $\mathcal{F}_{N}^{(2)}(\mathbf k_1,\mathbf k_2,\eta)$  is the Newtonian contribution, corresponding to the particular solution in Eq.~(\ref{partplushom})
\begin{equation}
\mathcal{F}_{N}^{(2)}(\mathbf k_1,\mathbf k_2,\eta)=\Bigg\{\frac{5}{7}+\frac{2}{7}\frac{(\mathbf k_1\cdot\mathbf k_2)^2}{k_1^2k_2^2}+\frac{\mathbf k_1\cdot\mathbf k_2(k_1^2+k_2^2)}{2k_1^2k_2^2}\Bigg\},
\label{Kernel2N}
\end{equation}

\noindent the relativistic corrections $\mathcal{F}_{R}^{(2)}(\mathbf k_1,\mathbf k_2,\eta)$, obtained from Eqs.~(\ref{delta1N}) and (\ref{Reldelta2}), in Fourier space are given by 
\begin{equation}
\mathcal{F}_{R}^{(2)}(\mathbf k_1,\mathbf k_2,\eta)=3\mathcal{H}^2\Bigg\{\left(f_{\rm NL}-\frac{5}{3}\right)\frac{k_1^2+k_2^2}{2k_1^2k_2^2}+\left(f_{\rm NL}+\frac{5}{12}\right)\frac{\mathbf k_1\cdot \mathbf k_2}{k_1^2k_2^2}\Bigg\}.
\label{Kernel2R}
\end{equation}

Similarly, the third order density contrast is defined as 

\begin{align}
\label{delta3N}
\frac{\delta^{(3)}(\mathbf k,\eta)}{6}=\int\frac{d^3\mathbf k_1d^3\mathbf k_2d^3\mathbf k_3}{(2\pi)^6}&\delta_D(\mathbf k-\mathbf k_1-\mathbf k_2-\mathbf k_3 )\mathcal{F}^{(3)}(\mathbf k_1,\mathbf k_2,\mathbf k_3,\eta)\\
\times&\delta^{(1)}(\mathbf k_1,\eta)\delta^{(1)}(\mathbf k_2,\eta)\delta^{(1)}({\mathbf k_3},\eta),
\nonumber
\end{align}

\noindent where the kernel $\mathcal{F}^{(3)}(\mathbf k_1,\mathbf k_2,\mathbf k_3,\eta)$ is also composed by Newtonian and relativistic contributions

\begin{equation}
\mathcal{F}^{(3)}(\mathbf k_1,\mathbf k_2,\mathbf k_3,\eta)=\mathcal{F}_{N}^{(3)}(\mathbf k_1,\mathbf k_2,\mathbf k_3,\eta)+\mathcal{F}_{R}^{(3)}(\mathbf k_1,\mathbf k_2,\mathbf k_3,\eta),
\label{Kernels3}
\end{equation}

\noindent with the third order Newtonian kernel given by \cite{Djeong}
\begin{align}
 \label{Kernel3N}
\nonumber
\mathcal{F}^{(3)}_{N}(\mathbf k_1,\mathbf k_2,\mathbf k_3,\eta)=&\frac{2k^2}{54}\left[\frac{\mathbf k_1\cdot\mathbf k_{23}}{k_1^2k_{23}^2}G_2^{(s)}(\mathbf k_2,\mathbf k_3)+(2\ \rm cyclic)\right]\\
           +&\frac{7}{54}\mathbf k\cdot\left[\frac{\mathbf k_{12}}{k_{12}^2}G_2^{(s)}(\mathbf k_1,\mathbf k_2)+(2\ \rm cyclic)\right]\\ 
\nonumber           
            +&\frac{7}{54}\mathbf k\cdot\left[\frac{\mathbf k_{1}}{k_{1}^2}F_2^{(s)}(\mathbf k_2,\mathbf k_3)+(2\ \rm cyclic)\right],\\
            \nonumber
\end{align}
\noindent and the relativistic contribution from Eqs.~($\ref{delta1N}$) and ($\ref{Reldelta3}$)
\begin{align}
\nonumber
\mathcal{F}_{R}^{(3)}(\mathbf k_1,\mathbf k_2,\mathbf k_3,\eta)=\frac{27}{2}\mathcal{H}^4\Bigg[-&\frac{\mathbf k_1\cdot \mathbf k_2+\mathbf k_1\cdot\mathbf k_3+\mathbf k_2\cdot\mathbf k_3}{k_1^2k_2^2k_3^2}
\bigg(-g_{\rm NL}+\frac{5}{9}f_{\rm NL}+\frac{25}{54}\bigg)\\
-&\frac{1}{6}\frac{ k_1^2+k_2^2+k_3^2}{k_1^2k_2^2k_3^2}\bigg(-g_{\rm NL}+\frac{10}{3}f_{\rm NL}-\frac{50}{27}\bigg)\Bigg].
\label{Kernel3R}
\end{align}

\section{One-loop power spectrum}
\label{sec:One-loop}
The $n^{th}$ order contribution to the density power spectrum $P^{(n)}(\mathbf{k}, \eta)$ \cite{Bernardeau:2001qr} is defined as, 
\begin{equation}
(2\pi)^3\delta^{D}(\mathbf k+\mathbf k')P^{(n)}(\mathbf k,\eta)=\sum_{m=1}^{2n-1}\frac{1}{m!{(2n-m)!}}\langle \delta_m(\mathbf k,\eta)\delta_{2n-m}(\mathbf k',\eta)\rangle.
\end{equation}
\noindent From this expression we find the first order power spectrum $P^{(1,1)}(k,\eta)$, also known as the tree-level power spectrum, corresponding to the linear power spectrum $P_{L}(k,\eta)$. Writing all the contributions up to second order ($n=2$) for the density power spectrum we obtain \cite{Scoccimarro:1996se}:
\begin{equation}
P(k,\eta)=P_L(k,\eta)+2P^{(1,3)}(k,\eta)+P^{(2,2)}(k,\eta),
\label{Oneloop}
\end{equation}

\noindent where  $P^{(1,3)}(k,\eta)$ and $P^{(2,2)}(k,\eta)$ corrections are known as the one-loop corrections to the density power spectrum. Since $\delta^{(1)}$ is a Gaussian field, correlations of the order $P^{(1,2)}(k,\eta)$ are null (in contrast with the expansions presented in e.g.~\cite{Taruya:2008pg,Desjacques:2008vf}).
 We use the solutions for the density contrast presented in Eqs.~(\ref{delta2N}) and (\ref{delta3N}) to calculate the one-loop density power spectrum.

\subsection{Second order density power spectrum correction $P^{(2,2)}(k,\eta)$}

The second order contribution to the density power spectrum $P^{(2,2)}(k,\eta)$ is defined as 
\begin{equation}
P^{(2,2)}(k,\eta)=2\int\frac{d^3q}{(2\pi)^3}P_L(q,\eta)P_L(|\mathbf k-\mathbf q|,\eta)[\mathcal{F}^{(2)}(\mathbf q,\mathbf k-\mathbf q,\eta)]^2.
\end{equation}

\noindent After substituting the expressions for $\mathcal{F}^{(2)}(\mathbf q,\mathbf k-\mathbf q,\eta)$ defined in Eq.~(\ref{Kernels2})  and using the following variable transformation \cite{Makino:1991rp} 

 \begin{equation}
 x=\frac{\mathbf k\cdot\mathbf q}{|\mathbf k||\mathbf q|}=\cos{\theta}, \quad r=\frac{|\mathbf q|}{|\mathbf k|},
 \label{variablechange}
 \end{equation} 
 
\noindent we can write the total second order power spectrum correction $P^{(2,2)}(k,\eta)$ as a sum of a Newtonian density power spectrum $P_{NN}^{(2,2)}(k,\eta)$, a cross term  $P_C^{(2,2)}(k,\eta)$ that includes Newtonian and relativistic terms, and a purely relativistic term $P_{RR}^{(2,2)}(k,\eta)$
 
\begin{equation}
P^{(2,2)}(k,\eta)=P_{NN}^{(2,2)}(k,\eta)+P_C^{(2,2)}(k,\eta)+P_{RR}^{(2,2)}(k,\eta).
\end{equation}
 
 \noindent Altogether this is 
 \begin{align}
 \nonumber
 P^{(2,2)}(k,\eta)=&\frac{k^3}{2\pi^2}\int_{0}^{\infty}r^2dr  P_L(kr,\eta)\int_{-1}^{1}dxP_L(k\sqrt{1+r^2-2rx},\eta)\\
  \times&\Bigg\{  \bigg[ \frac{3r+7x-10rx^2}{14r(1+r^2-2rx)}\bigg]^2 \label{integral22}\\
  \nonumber
+& \mathcal{H}^2\frac{(6f_{\rm NL}-10-25r(r-x))(3r+7x-10rx^2)}{28k^2r^3(1+r^2-2rx)^2}\\
 +&  \bigg[{\mathcal{H}^2}\frac{6f_{\rm NL}-10-25r^2+25rx}{4r^2k^2(1-2rx+r^2)}\bigg]^2\Bigg\},
 \nonumber
 \end{align}

\noindent where the first and second lines correspond to $P_{NN}^{(2,2)}(k,\eta)$, while the  third and fourth lines correspond to  $P_C^{(2,2)}(k,\eta)$ and $P_{RR}^{(2,2)}(k,\eta)$ respectively.

\subsection{Second order density power spectrum correction $P^{(1,3)}(k,\eta)$}
The second order contribution $P^{(1,3)}(k,\eta)$ is defined as 

\begin{equation}
P^{(1,3)}(k,\eta)=3\mathcal{F}^{(1)}(\mathbf k)P_L(k,\eta)\int\frac{d^3q}{(2\pi)^3}P_L(q,\eta)\mathcal{F}^{(3)}(\mathbf k,\mathbf q,-\mathbf q,\eta),
\end{equation}
 
\noindent where $\mathcal{F}^{(3)}(\mathbf k,\mathbf q,-\mathbf q,\eta)$ is defined by Eq.~(\ref{Kernels3}) and is written in terms of the variables defined in Eq.~(\ref{variablechange}).  For the total second order contribution $P^{(1,3)}(k,\eta)$ we have the sum of a Newtonian contribution $P_{NN}^{(1,3)}(k,\eta)$ and a relativistic contribution $P_{RR}^{(1,3)}(k,\eta)$, where  we do not have cross terms as we do not have relativistic corrections in the first order kernel $\mathcal{F}^{(1)}(\mathbf k)$
 
\begin{equation}
P^{(1,3)}(k,\eta)=P_{NN}^{(1,3)}(k,\eta)+P_{RR}^{(1,3)}(k,\eta),
\end{equation}

\noindent using the change of variables in (\ref{variablechange}) and integrating over the variable $x$, we obtain
\begin{align}
\nonumber
P^{(1,3)}(k,\eta)=\frac{k^3}{4\pi^2}P_L(k,\eta)\int_{0}^{\infty} drP_L(kr,\eta)&\Bigg\{ \frac{1}{504}\bigg[\frac{12}{r^2}-{158}+100r^2-42r^4\\
+&\frac{3}{r^3}(r^2-1)^3(7r^2+2)\ln{\left(\frac{r+1}{|r-1|}\right)}\bigg] \label{integral13}\\
\nonumber
+&\frac{81}{k^4}{\mathcal{H}^4}
\bigg[\Big(-g_{\rm NL}+\frac{5}{9}f_{\rm NL}+\frac{25}{54}\Big)\\
\nonumber
-&\frac{1+2r^2}{6r^2}\Big(-g_{\rm NL}+\frac{10}{3}f_{\rm NL}-\frac{50}{27}\Big) \bigg]\Bigg\},
\end{align}

\noindent where the first and second line correspond to $P_{NN}^{(1,3)}(k,\eta)$ and third and fourth line to $P_{RR}^{(1,3)}(k,\eta)$.

We obtain numerical solutions for the different contributions to the density power spectrum presented in this section. All our integrations use as an input a linear power spectrum generated with the Boltzmann solver CLASS \cite{Lesgourgues:2011re}, assuming a flat $\Lambda$CDM cosmology given by the Planck collaboration \cite{Ade:2015xua} with a sharp cut-off in $P_L(k)$ at $k=10^{-5}h{\rm Mpc}^{-1}$ due to the infrared behaviour of the purely relativistic terms (see the Appendix \ref{IRL}). To test the convergence of the numerical integration of the density power spectrum, we have computed these integrals with the Mathematica package and with a Python script independently.

In this way, the total Newtonian one-loop power spectrum is given as usual by
\begin{equation}
P_{NN}(k,\eta)=P_L(k,\eta)+2P_{NN}^{(1,3)}(k,\eta)+P_{NN}^{(2,2)}(k,\eta).
\end{equation}

\begin{figure}[tbp]
\centering
\includegraphics[width=110mm]{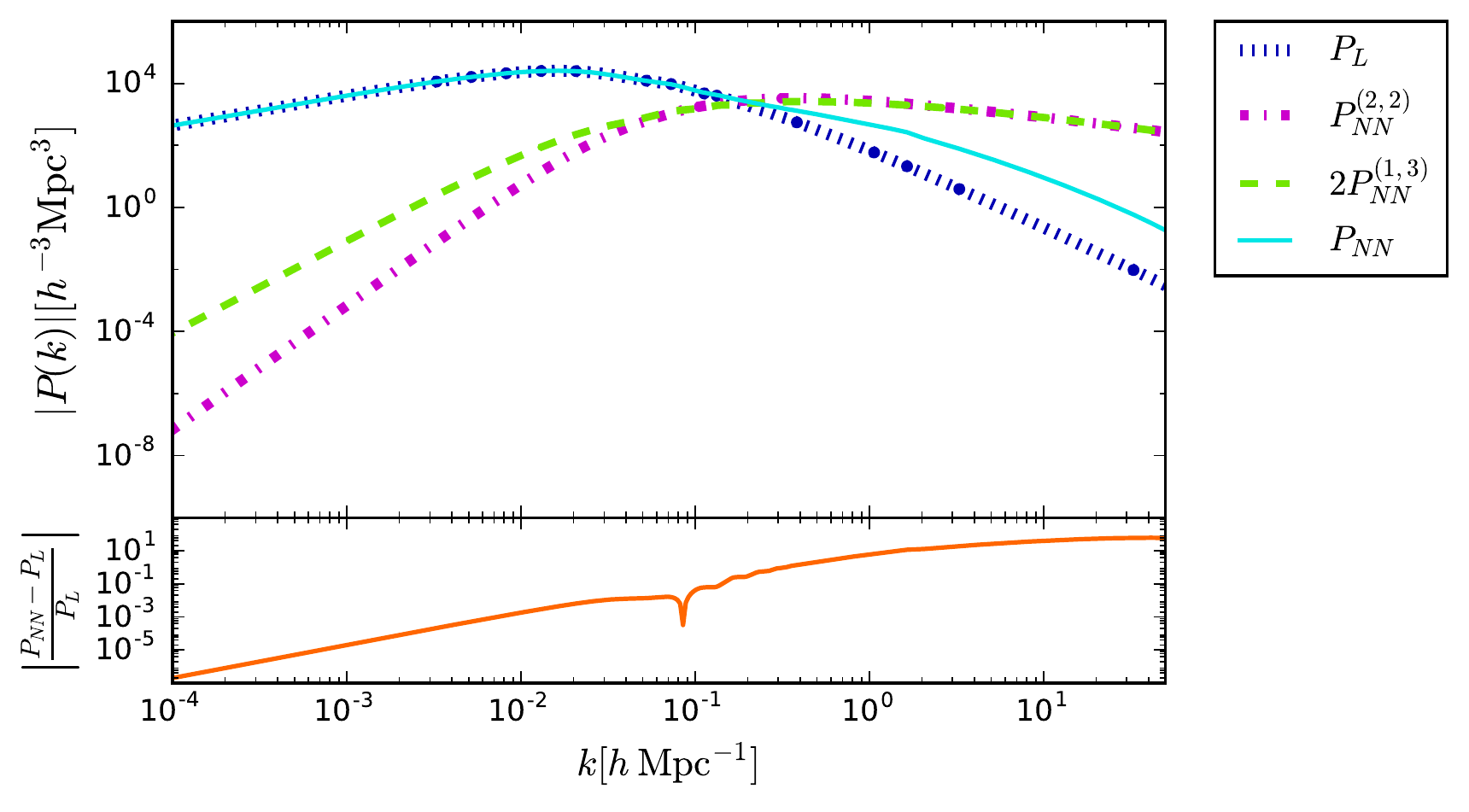}
\caption{Upper panel: The total Newtonian one-loop power spectrum $P_{NN}$ and the individual second order density power spectrum contributions to the total Newtonian one-loop power spectrum $P^{(2,2)}_{NN}$ and $P^{(1,3)}_{NN}$, at redshift $z=0$. Bottom panel: The relative difference of the Newtonian one-loop power spectrum with respect to the linear power spectrum normalised with the linear power spectrum. 
 }
\label{NewtonOne}
\end{figure}

In Figure \ref{NewtonOne} we present the Newtonian standard perturbation theory results, showing the second order Newtonian contributions to the one-loop power spectrum, $P_{NN}^{(2,2)}$ and $P_{NN}^{(1,3)}$, along with the total Newtonian one-loop power spectrum $P_{NN}$, for comparison we also plot the linear power spectrum $P_L$ in all the figures presented. The relative difference of the Newtonian one-loop power spectrum with respect to the linear power spectrum is also shown. The Newtonian contributions show a relevant effect only for the small scales.

The total relativistic one-loop power spectrum is defined as 
\begin{equation}
P_{RR}(k,\eta)=P_L(k,\eta)+2P_{RR}^{(1,3)}(k,\eta)+P_{RR}^{(2,2)}(k,\eta).
\end{equation}

\begin{figure}[tbp]
\centering
\includegraphics[width=110mm]{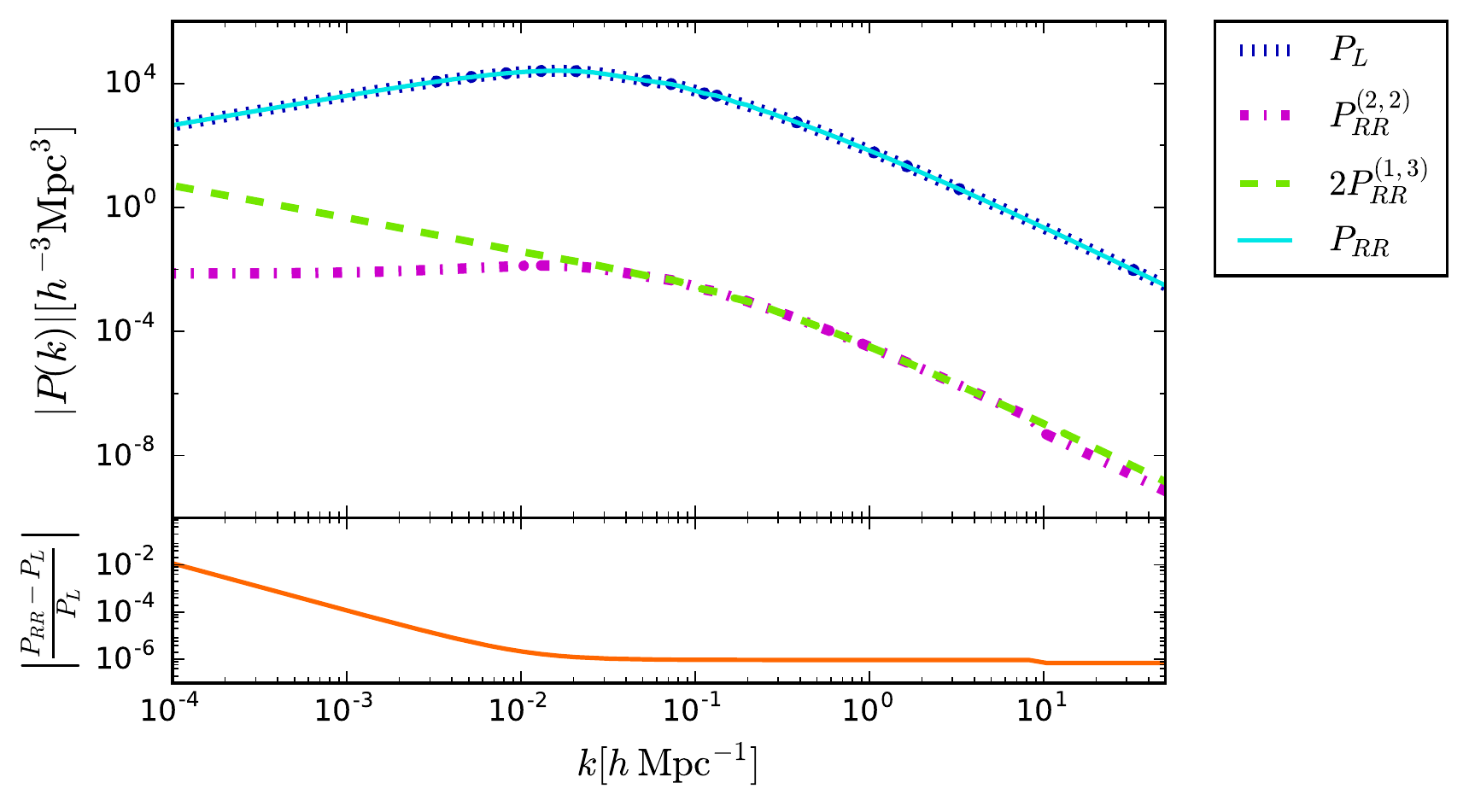}
\caption{Upper panel: The total relativistic one-loop power spectrum $P_{RR}$ and the individual second order density power spectrum contributions to the total relativistic one-loop power spectrum $P^{(2,2)}_{RR}$ and $P^{(1,3)}_{RR}$ for $f_{\rm NL}=g_{\rm NL}=0$, at redshift $z=0$. Bottom panel: The relative difference of the relativistic one-loop power spectrum  with respect to the linear power spectrum normalised with the linear power spectrum. }
\label{GROne}
\end{figure}

In Figure \ref{GROne} we present the relativistic results, we show the relativistic contributions to the one-loop power spectrum coming from, $P_{RR}^{(2,2)}$ and $P_{RR}^{(1,3)}$, along with the total relativistic one-loop power spectrum $P_{RR}$, in this Figure we consider the case in where $f_{\rm NL}=g_{\rm NL}=0.$ The relative difference of the relativistic one-loop power spectrum  with respect to the linear power spectrum is also shown. We note that relativistic one-loop power spectrum  corrections are relevant in the large scales, the relativistic contributions are subdominant in smaller scales. 

Finally, the total one-loop power spectrum defined in Eq.~(\ref{Oneloop}) reads as
\begin{equation}
P_{RN}(k,\eta)=P_L(k,\eta)+2P^{(1,3)}(k,\eta)+P^{(2,2)}(k,\eta).
\label{Oneloopexp}
\end{equation}

\begin{figure}[tbp]
\centering
\includegraphics[width=110mm]{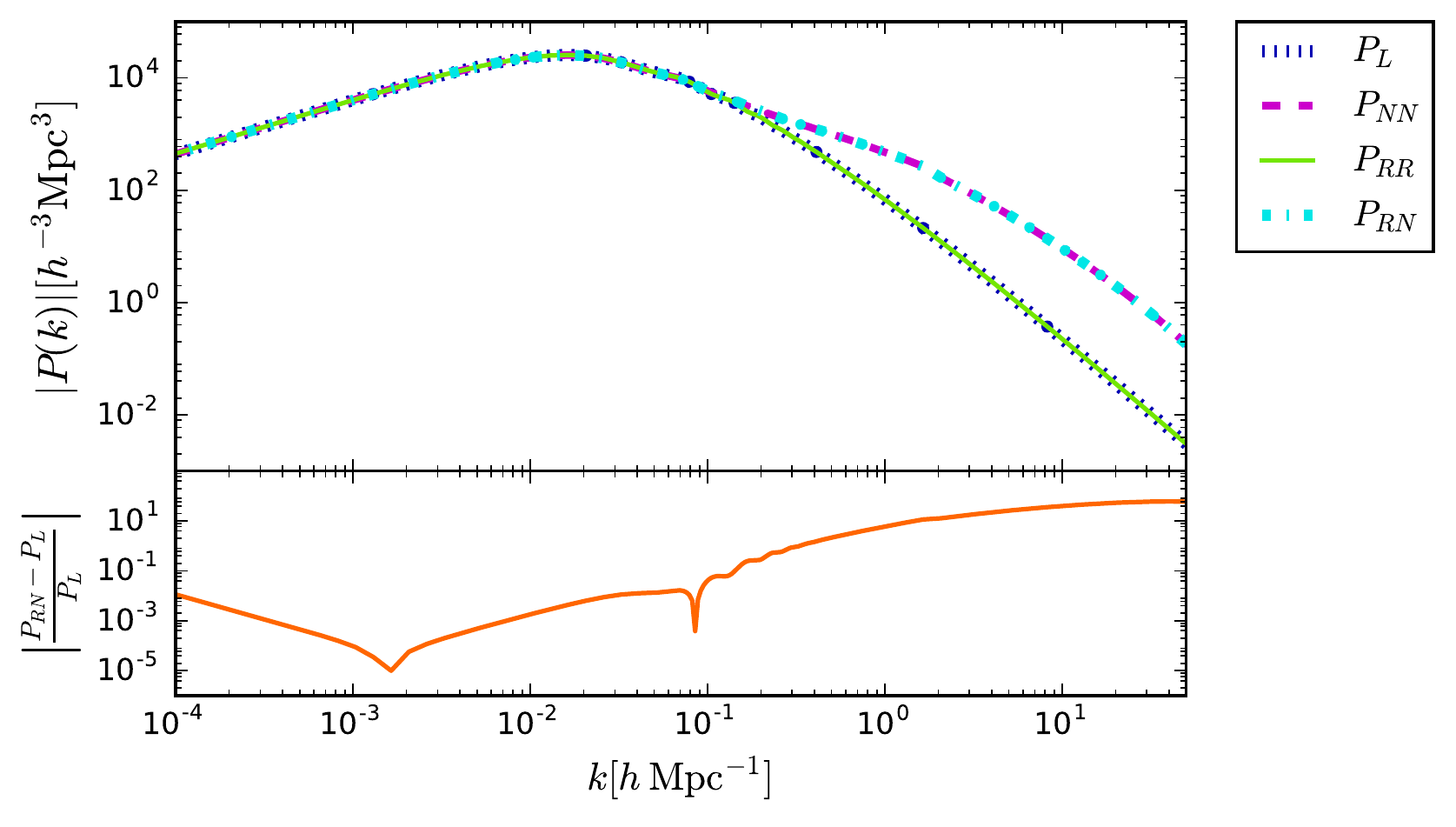}
\caption{Upper panel: Total Newtonian one-loop power spectrum $P_{NN}$ along with the total relativistic one-loop power spectrum $P_{RR}$ and total one-loop power spectrum $P_{RN}$, with $f_{\rm NL}=g_{\rm NL}=0$, at redshift $z=0$. Bottom panel: Relative difference of the total one-loop power spectrum  with respect to the linear power spectrum normalised with the linear power spectrum.}
\label{TotalOne}
\end{figure}

In Figure \ref{TotalOne}  we present a comparison of the total Newtonian one-loop power spectrum $P_{NN}$, the total relativistic one-loop power spectrum $P_{RR}$, along with the total one-loop power spectrum $P_{RN}$, in this Figure we consider the case with no primordial non-Gaussianity $f_{\rm NL}=g_{\rm NL}=0.$ The difference of the total one-loop power spectrum $P_{RN}$ respect to the linear power spectrum $P_L$ lies in the large scales is due to the relativistic corrections, whereas the difference in the small scales is given purely by the Newtonian contributions.

\begin{figure}[tbp]
\centering
\includegraphics[width=150mm]{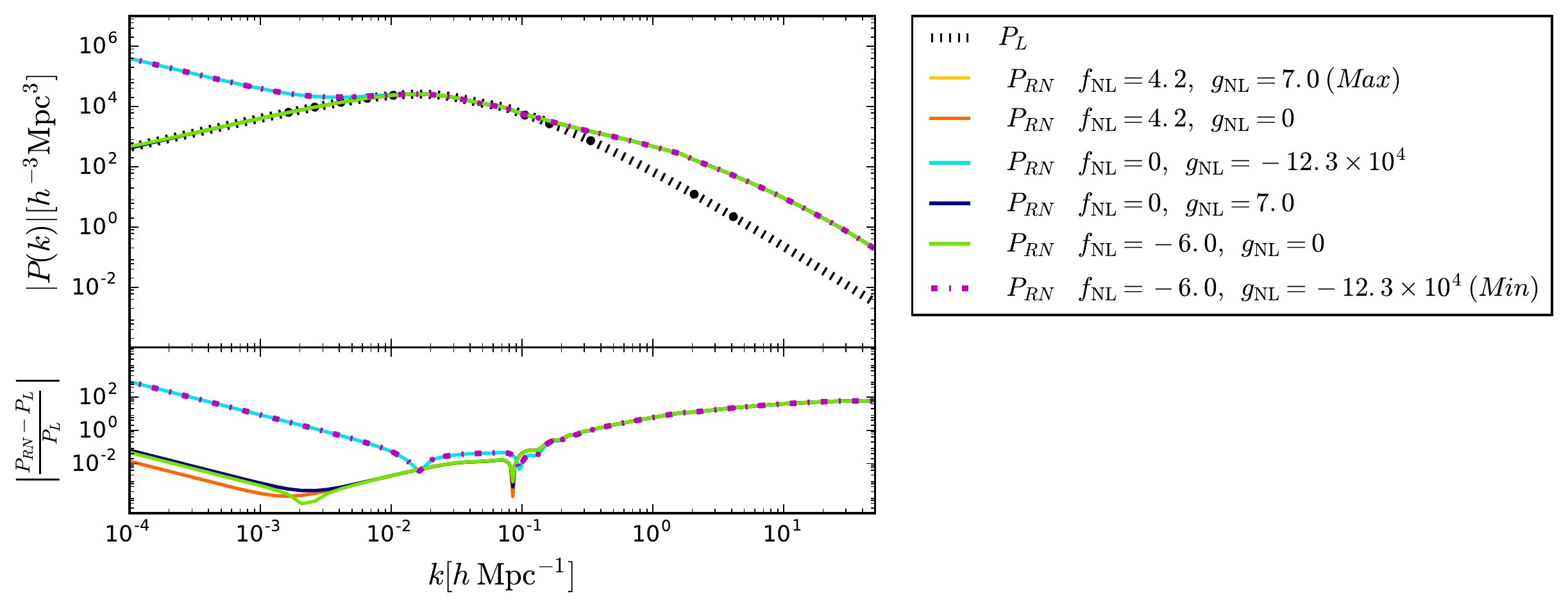}
\caption{Upper panel: Total one-loop power spectrum $P_{RN}$, at redshift $z=0$, for different limiting   values of $f_{\rm NL}$ and $g_{\rm NL}$ reported by Planck\cite{Akrami:2019izv}. Bottom panel: Relative difference of the relativistic one-loop power spectrum with respect to the linear power spectrum  normalised with the linear power spectrum.}
\label{Combinations}
\end{figure}

In Figure \ref{Combinations} we present the total one-loop power spectrum $P_{RN}$, using different combinations of values of $f_{\rm NL}$ and $g_{\rm NL}$ reported in by the  Planck collaboration in Ref.~\cite{Akrami:2019izv}. The current constraints are given by $f^{\rm local}_{\rm NL}=-0.9\pm5.1$ and $g^{\rm local}_{\rm NL}=-5.8\pm6.5\times10^{4}$. For $f_{\rm NL}$ we use the minimum and maximum values allowed by Planck i.e. $f_{\rm NL}=-6.0$ and $f_{\rm NL}=4.2$. In the case of $g_{\rm NL}$, we use the minimum value allowed by Planck i.e. $g_{\rm NL}=-12.3\times10^4$, however the maximum value of $g_{\rm NL}$ that we can use is $\sim 7$ as higher values, although allowed by the Planck collaboration \cite{Akrami:2019izv}, give negative, non-perturbative contributions to the density power spectrum on large scales. These values for $g_{\rm NL}$ and $f_{\rm NL}$ were chosen to show which values of $f_{\rm NL}$ and $g_{\rm NL}$ have a more significant contribution to the one-loop power spectrum. The relative difference with respect to the linear power spectrum shows that the largest corrections to the power spectrum in the large scales are present when $g_{\rm NL}$ takes its minimum value, being this the dominant correction term as is not affected by the chosen value of $f_{\rm NL}$. On the other hand, larger values of $g_{\rm NL}$ present a similar behaviour for the different combinations with $f_{\rm NL}$, having a small relative difference with respect to the linear power spectrum in comparison to the corrections given by minimum values of $g_{\rm NL}$.

\begin{figure}[tbp]
\centering
\includegraphics[width=150mm]{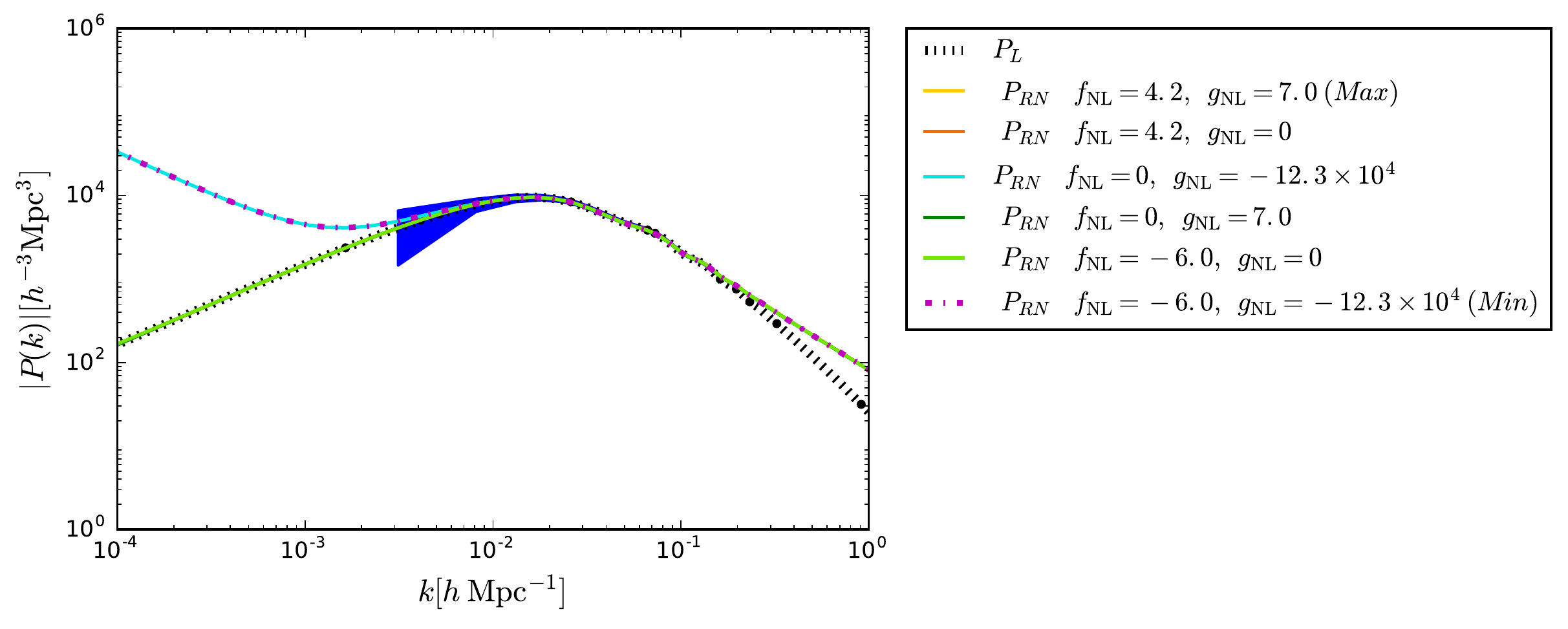}
\caption{Total one-loop power spectrum $P_{RN}$, at redshift $z=1$, for different limiting  values of $f_{\rm NL}$ and $g_{\rm NL}$ reported by Planck \cite{Akrami:2019izv}. The blue shaded area corresponds to the measurement error of a typical Stage-IV-like survey redshift bin with $\Delta z = 0.2$, as detailed in the main text. We also used a $k$-binning $\Delta k = 0.006 \, h{\rm Mpc^{-1}}$.}
\label{Combinationsz1}
\end{figure}

In Figure \ref{Combinationsz1} we present the same set of total one-loop power spectrum $P_{RN}$ plots as in Figure \ref{Combinations} but at a redshift $z=1$. In addition to the density power spectrum $P_{RN}$ we also present in the blue shaded area the measurement errors assuming a cosmic variance limited Stage-IV galaxy survey like DESI \cite{Aghamousa:2016zmz}, Euclid \cite{Blanchard:2019oqi}, or LSST \cite{Chisari:2018vrw}. More specifically, we have assumed a sky area of $15,000 \, {\rm deg}^2$ at $z=1$ with bin width $\Delta z = 0.2$. These numbers correspond to typical specifications of such surveys used in recent forecast and model validation studies at $z=1$ (see e.g. \cite{Markovic:2019sva}). Note however that the measurement errors would decrease if we chose a wider redshift bin given the large \emph{total} redshift coverage of Stage IV surveys. Similarly, we have defined the largest measurable scale as $k_{\rm min} \simeq 2\pi / V^{1/3}_{\rm bin} = 0.003 \, h{\rm Mpc^{-1}}$, where $V_{\rm bin}$ is the volume corresponding to $\Delta z = 0.2$; this volume would increase if we were to consider a wider redshift bin, allowing us to reach larger scales. Note that the minimum values of $f_{\rm NL}$ and $g_{\rm NL}$ show the largest impact at the largest measured scales of the upcoming experiments, forecasting a detectability of PNG for values of $g_{\rm NL}$ or $f_{\rm NL}$.

\section{Tree-level bispectrum}
\label{sec:Treebis}
For completeness we calculate the tree-level bispectrum, which is  defined as 
\begin{equation}
B(k_1,k_2,k_3,\eta)\equiv2P_L(k_1,\eta)P_L(k_2,\eta)\mathcal{F}^{(2)}(\mathbf k_1,\mathbf k_2)+\rm{(2\enspace cyclic)},
\label{bispe}
\end{equation}
\noindent the components to calculate the bispectrum at tree-level are already given in Eqs.~(\ref{Kernel2N}) and (\ref{Kernel2R}).
We define the Newtonian tree-level bispectrum $B_{NN}$ as

\begin{equation}
B_{NN}(k_1,k_2,k_3,\eta)\equiv2P_L(k_1,\eta)P_L(k_2,\eta)\mathcal{F}_{N}^{(2)}(\mathbf k_1,\mathbf k_2)+\rm{(2\enspace cyclic)},
\end{equation}

\noindent and the  relativistic tree-level bispectrum as
\begin{equation}
B_{RR}(k_1,k_2,k_3,\eta)\equiv2P_L(k_1,\eta)P_L(k_2,\eta)\mathcal{F}_{R}^{(2)}(\mathbf k_1,\mathbf k_2)+\rm{(2\enspace cyclic)},
\end{equation}

the total tree-level bispectrum $B_{RN}$,  is defined by Eq.~(\ref{bispe}), where $\mathcal{F}^{(2)}(\mathbf{k}_1,\mathbf{k}_2,\eta)$ is given in Eq.~(\ref{Kernels2}).

\begin{figure}
\begin{subfigure}{\linewidth}
\centering
\includegraphics[width=75mm]{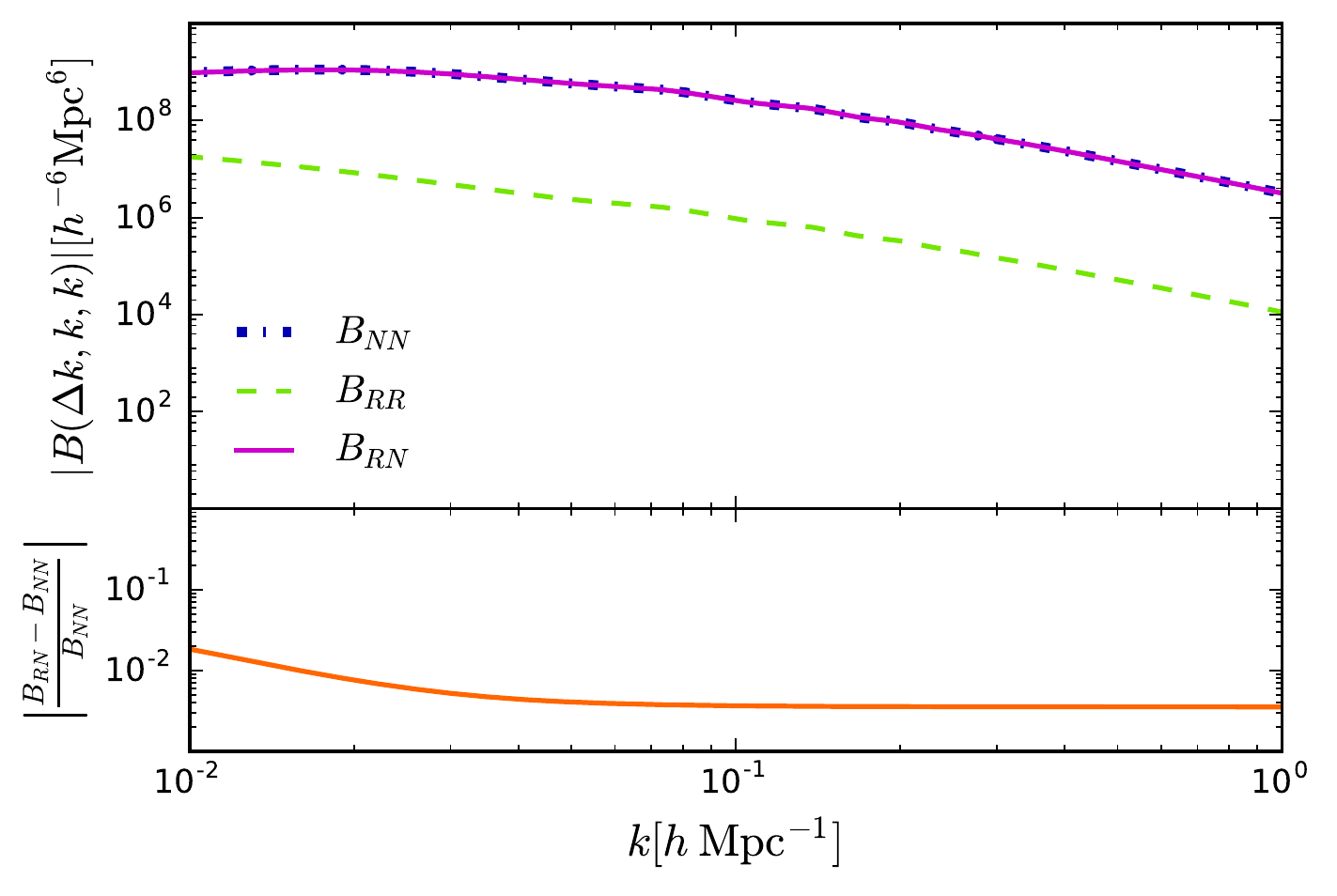}
\caption{}
\label{Bi1}
\end{subfigure}\\[1ex]
\begin{subfigure}{.5\linewidth}
\centering
\includegraphics[width=75mm]{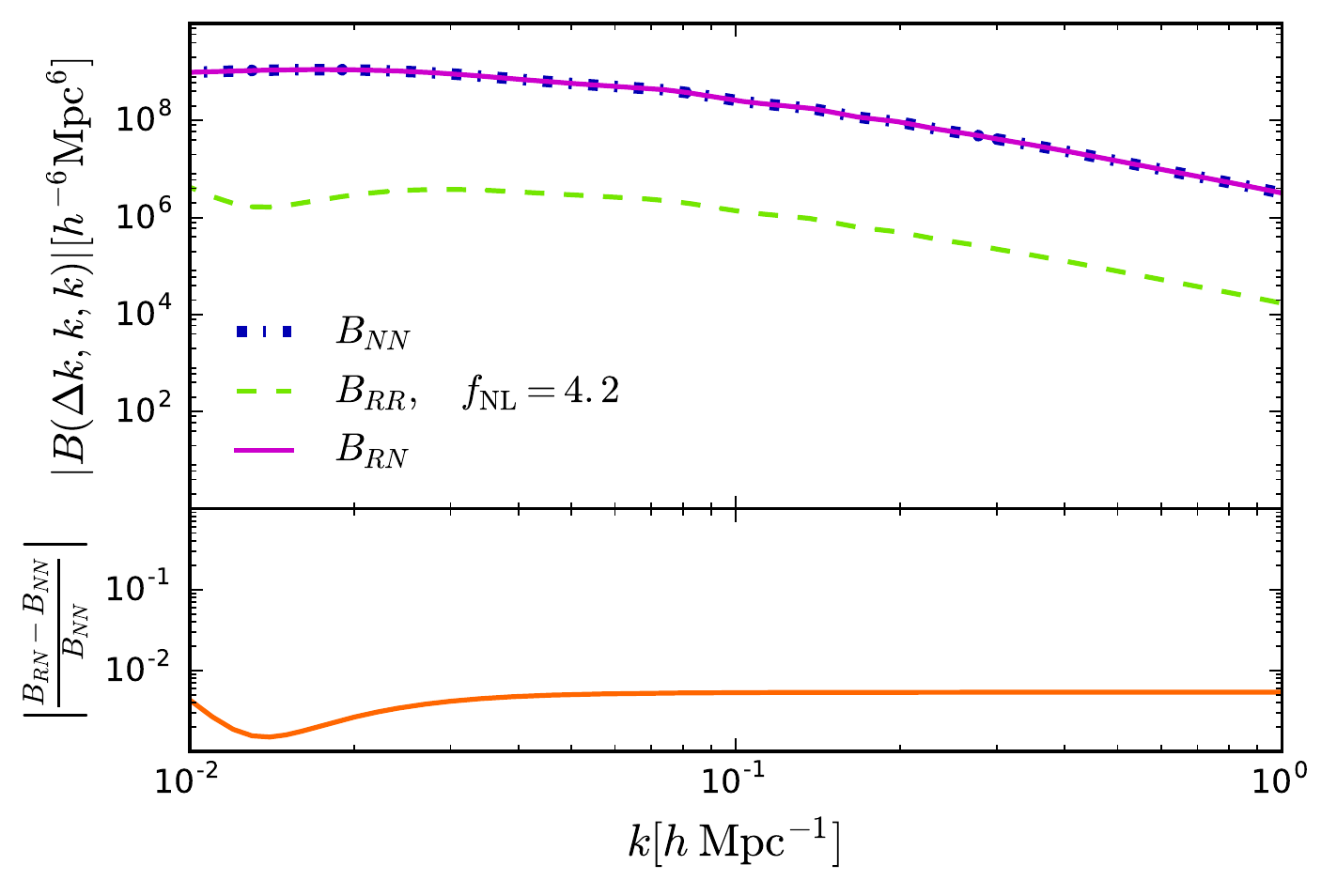}
\caption{}
\label{Bi2}
\end{subfigure}
\begin{subfigure}{.5\linewidth}
\centering
\includegraphics[width=75mm]{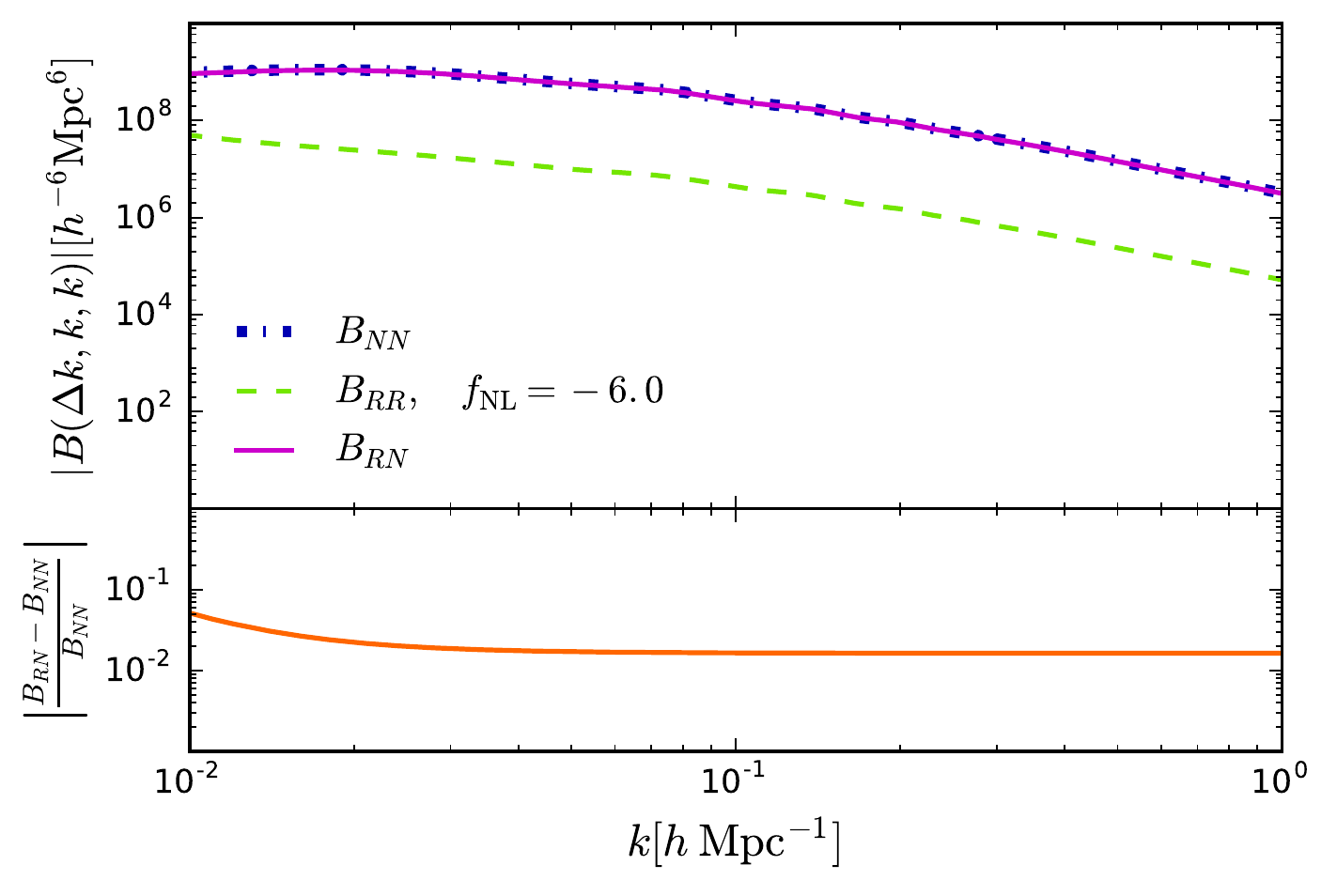}
\caption{}
\label{Bi3}
\end{subfigure}
\caption{Upper panels: Comparison of the Newtonian, relativistic and total tree-level bispectrum corrections in the squeezed limit with $\Delta k=0.013\:h\rm{Mpc}^{-1}$ for (a) $f_{\rm NL}=0$, (b) $f_{\rm NL}=4.2$ and (c) $f_{\rm NL}=-6.0$, at redshift $z=0$. Bottom panels: Relative difference of the total tree-level bispectrum with respect to the Newtonian tree-level bispectrum normalised with the Newtonian tree-level.}
\label{Bispectrum}
\end{figure}

In Figure \ref{Bispectrum} we present a comparison of the Newtonian tree-level bispectrum $B_{NN}$, the  relativistic tree-level bispectrum given by $B_{RR}$ and the total tree-level bispectrum $B_{RN}$, all in the squeezed limit, with $\Delta k=0.013\:h\rm{Mpc}^{-1}$ when $f_{\rm NL}=0$  and for the limiting values of $f_{\rm NL}$ given by Ref.~\cite{Akrami:2019izv}, the relative difference of the total tree-level bispectrum  with respect to the Newtonian bispectrum is shown in the bottom panels. The relativistic corrections at this level are subdominant with respect to the Newtonian tree-level bispectrum.

\section{Discussion}
\label{sec:conclusions}

We calculated purely general relativistic corrections to the density power spectrum at one-loop. For the synchronous-comoving gauge the primordial non-Gaussianity of the local type can be added naturally and we have also computed the contribution of these parameters. 
The modifications that relativistic contributions bring to the density power spectrum are below  $0.01\%$ except at very large scales  where we find a $1\%$ pure relativistic contribution (see Figure~\ref{GROne}). 
On the other hand, the primordial non-Gaussianity values allowed by the latest Cosmic Microwave Background observations in the local configuration yield significant contributions mostly from the $g_{\rm NL}$ parameter (see Figure~\ref{Combinationsz1}).

The relativistic terms contributing to the higher order amplitude of the density contrast have been derived from a long-wavelength approximation and do not account for effects at all scales. However, it is expected that at small scales the weak field and therefore the Newtonian regime describe best the matter structure. As mentioned above, it is precisely at the large scales where primordial non-Gaussianity contributes to the density power spectrum. Therefore, the formalism employed here to derive relativistic contributions is naturally extended to include the dominant PNG contributions to the density contrast and its polispectra.

The actual corrections from GR to the non-linear bias of galaxies seem to remove the GR effects presented here, however the form of the volume distortions reproduce the form of those expressed in our non-linear prescriptions for the density contrast---at least at second order \cite{Umeh:2019qyd} but decoupling large and short scales---.  Yet the local primordial non-Gaussianity terms cannot be removed by local coordinate transformations \cite{Umeh:2019jqg}, and terms with such factors are precisely what dominates the signal in the one-loop spectra. It is also important to mention that the GR effects removed by the coordinate transformation are carried at second order and the third order effects described here might survive the coordinate changes. The details of adapting the expressions coming from the volume distortions as galaxies evolve from an initial time, and the precise consequences for a non-linear galaxy power spectrum are left for future work.

Our results show that pure relativistic corrections $P_{RR}$ have a too small contribution at too large scales to be observed in the present or future large scale structure probes. On the other hand, the primordial non-Gaussianity contributions, corresponding to values within the 1-$\sigma$ amplitudes of $g_{\rm NL}$ allowed by Planck \cite{Akrami:2019izv}, yield a significant contribution to $\delta^{(3)}$, and to the one-loop power spectrum observable in the next generation of galaxy surveys. While the deviations from the linear prescription lie within the cosmic variance errors, it may be possible to probe these values through cross-correlations of the future surveys with the measurements of anisotropies in the Cosmic Microwave Background. We shall explore the implications of this effect in order to constrain primordial non-Gaussianity through this and other methods in a future work.

\acknowledgments 

The authors thank Pedro Carrilho and Anthony Lewis for their useful
comments. We thank Alkistis Pourtsidou for her useful comments and
help with Figure \ref{Combinationsz1}. RMC acknowledges support of a
studentship funded by Queen Mary University of London and CONACyT
grant No. 661285. The authors acknowledge sponsorship from CONACyT
through grant CB-2016-282569. KAM is funded in part by STFC grant
ST/P000592/1.

\appendix
\section{Infrared limits}\label{IRL}

The infrared (IR) contributions of the one-loop integrals (\ref{integral22}) and (\ref{integral13})
can be computed as the part of the integral from $r=0$ to a small value $\epsilon$. With this consideration, the one-loop power spectrum can be written as:
\begin{eqnarray}
P^{(2,2)}(k,\eta) &=& \frac{k^3}{2\pi^2}\left( \int_0^\epsilon + \int_\epsilon^\infty \right)
r^2 dr 
P_L(kr,\eta)\int_{-1}^{1}dxP_L(k\sqrt{1+r^2-2rx},\eta) \nonumber\\
&&  \times \left[ (\mathcal{F}^{(2)}_N)^2+2 \mathcal{F}^{(2)}_N\mathcal{F}^{(2)}_R+(\mathcal{F}^{(2)}_R)^2 \right]
\,, \\
P^{(1,3)}(k,\eta) &=& \frac{k^3}{4\pi^2}P_L(k,\eta) \left( \int_0^\epsilon + \int_\epsilon^\infty \right) dr  P_L(kr,\eta) \left[\mathcal{F}^{(3)}_N + \mathcal{F}^{(3)}_R \right] \,,
\end{eqnarray}
where the integrals in $r$ have been split between a possible divergent infrared contribution from 0 to $\epsilon$ and a finite contribution from $\epsilon$ to $\infty$ which, in the limit of $\epsilon \to 0$ will correspond to the Cauchy principal value of the integral.
Using $P_L(k,\eta)\propto k^{n_s}$ as $k\to 0$ the  infrared contributions  can be computed analytically, which we will write explicitly in the following expressions. Note that, in the cases where the integrals diverge we will write the expressions as the limit
\begin{equation}
    \int_0^\epsilon = \lim_{\delta \to 0} \int_\delta^\epsilon \,,
\end{equation}
in order to see divergence rate.
For the three different terms in $P^{(2,2)}$ we obtain the expressions:
\begin{eqnarray}
{}_{\rm{IR}}P^{(2,2)}_{NN} &=& \frac{k^3}{2\pi^2}P_L(k)P_L(k\epsilon) \frac{\epsilon}{3(n_s+1)} \,, \label{ir22}\\
{}_{\rm{IR}}P^{(2,2)}_{C} &=& \frac{\mathcal{H}^2k}{2\pi^2}P_L(k)P_L(k\epsilon) \frac{\epsilon}{n_s+1} 
  \left( \frac{27}{7}f_{\rm NL} - \frac{95}{42} \right)\,,\\
{}_{\rm{IR}}P^{(2,2)}_{RR}  &=&\frac{\mathcal{H}^4}{2\pi^2 k} P_L(k)P_L(k\epsilon)\left(  \frac{\epsilon A}{(n_s+1)}+ 
\frac{B}{\epsilon(n_s-1)}\lim_{\delta \to 0} \left( 1- \left( \frac{\delta}{\epsilon}\right)^{n_s-1}\right)
\right)  \,, \label{divergence22}
\end{eqnarray}
where
\begin{eqnarray}
A&=& 9f_{\rm NL}^2 - \frac{35}{2}f_{\rm NL}+\frac{725}{27} \,,\\
B&=& \frac{(3f_{\rm NL}-5)^2}{2}\,.
\end{eqnarray}
In these expressions the value of $\epsilon$ is small but fixed, meaning that the purely relativistic term diverges approximately as $\delta^{-0.03}$ for $\delta \to 0$. Meanwhile the infrared contributions to $P^{(1,3)}$ read:
\begin{eqnarray}
{}_{\rm{IR}}P^{(1,3)}_{NN} &=& - \frac{k^3}{4\pi^2} P_L(k)P_L(k\epsilon) \frac{\epsilon}{3(n_s+1)} \label{ir13} \,,\\
{}_{\rm{IR}}P^{(1,3)}_{RR}  &=&  \frac{\mathcal{H}^4}{4\pi^2 k} P_L(k)P_L(k\epsilon) \left(  \frac{\epsilon C}{(n_s+1)}+ 
\frac{D}{\epsilon(n_s-1)}\lim_{\delta \to 0} \left( 1- \left( \frac{\delta}{\epsilon}\right)^{n_s-1}\right)
\right) \,, \label{divergence13}
\end{eqnarray}
where
\begin{eqnarray}
C&=& 54 \left( -g_{\rm NL} - \frac{5}{6}f_{\rm NL} -\frac{175}{108} \right) \,,\\
D&=& -\frac{27}{2} \left( -g_{\rm NL} + \frac{10}{3}f_{\rm NL} -\frac{50}{27} \right) \,.
\end{eqnarray}
We see that the second term in (\ref{divergence13}) diverges at the same rate as (\ref{divergence22}).
For the purely Newtonian one loop contribution, the possible infrared problems in the different terms get solved as the combination $2{}_{\rm{IR}}P_{NN}^{(1,3)}(k,\eta)+{}_{\rm{IR}}P_{NN}^{(2,2)}(k,\eta)$ cancels out, as read from the expressions (\ref{ir22}) and (\ref{ir13}) (see Ref.~\cite{McEwen:2016fjn}). However for the relativistic term this does not happen as the expressions (\ref{divergence13}) and (\ref{divergence22}) do not cancel.

In order to obtain finite results for the relativistic one-loop contribution, we set a lower limit different from zero in the $r$ integrals. The fact that the divergence is very slow allows the results to not be very dependent on this limit, but only as $r_c^{-0.03}$. Moreover, as stated in Ref.~\cite{Castiblanco:2018qsd}, the observations have a minimum $k$ accessible to them, corresponding to their maximum observed scale.
Throughout this work we chose this limit to be in the parameter $q = kr$ as $q_c = 10^{-5}h{\rm Mpc^{-1}}$ which is close to the limit chosen in Ref.~\cite{Castiblanco:2018qsd} as $q_c = H_0 \approx 3\times 10^{-4}h{\rm Mpc^{-1}}$.

\bibliographystyle{JHEP.bst}

\bibliography{biblio}

\providecommand{\href}[2]{#2}\begingroup\raggedright\begin{thebibliography}{10}

\bibitem{Aghamousa:2016zmz}
{\scshape DESI} collaboration, \emph{{The DESI Experiment Part I:
  Science,Targeting, and Survey Design}},
  \href{https://arxiv.org/abs/1611.00036}{{\ttfamily 1611.00036}}.

\bibitem{Laureijs:2011gra}
{\scshape EUCLID} collaboration, \emph{{Euclid Definition Study Report}},
  \href{https://arxiv.org/abs/1110.3193}{{\ttfamily 1110.3193}}.

\bibitem{Makino:1991rp}
N.~Makino, M.~Sasaki and Y.~Suto, \emph{{Analytic approach to the perturbative
  expansion of nonlinear gravitational fluctuations in cosmological density and
  velocity fields}}, \href{https://doi.org/10.1103/PhysRevD.46.585}{\emph{Phys.
  Rev.} {\bfseries D46} (1992) 585}.

\bibitem{Jain:1993jh}
B.~Jain and E.~Bertschinger, \emph{{Second order power spectrum and nonlinear
  evolution at high redshift}},
  \href{https://doi.org/10.1086/174502}{\emph{Astrophys. J.} {\bfseries 431}
  (1994) 495} [\href{https://arxiv.org/abs/astro-ph/9311070}{{\ttfamily
  astro-ph/9311070}}].

\bibitem{Bernardeau:2001qr}
F.~Bernardeau, S.~Colombi, E.~Gazta{\~n}aga and R.~Scoccimarro, \emph{{Large
  scale structure of the universe and cosmological perturbation theory}},
  \href{https://doi.org/10.1016/S0370-1573(02)00135-7}{\emph{Phys. Rept.}
  {\bfseries 367} (2002) 1}
  [\href{https://arxiv.org/abs/astro-ph/0112551}{{\ttfamily
  astro-ph/0112551}}].

\bibitem{Bonvin:2011bg}
C.~Bonvin and R.~Durrer, \emph{{What galaxy surveys really measure}},
  \href{https://doi.org/10.1103/PhysRevD.84.063505}{\emph{Phys. Rev.}
  {\bfseries D84} (2011) 063505}
  [\href{https://arxiv.org/abs/1105.5280}{{\ttfamily 1105.5280}}].

\bibitem{Challinor:2011bk}
A.~Challinor and A.~Lewis, \emph{{The linear power spectrum of observed source
  number counts}},
  \href{https://doi.org/10.1103/PhysRevD.84.043516}{\emph{Phys. Rev.}
  {\bfseries D84} (2011) 043516}
  [\href{https://arxiv.org/abs/1105.5292}{{\ttfamily 1105.5292}}].

\bibitem{Castiblanco:2018qsd}
L.~Castiblanco, R.~Gannouji, J.~Noreña and C.~Stahl, \emph{{Relativistic
  cosmological large scale structures at one-loop}},
  \href{https://doi.org/10.1088/1475-7516/2019/07/030}{\emph{JCAP} {\bfseries
  1907} (2019) 030} [\href{https://arxiv.org/abs/1811.05452}{{\ttfamily
  1811.05452}}].

\bibitem{Bruni:2011ta}
M.~Bruni, R.~Crittenden, K.~Koyama, R.~Maartens, C.~Pitrou and D.~Wands,
  \emph{{Disentangling non-Gaussianity, bias and GR effects in the galaxy
  distribution}}, \href{https://doi.org/10.1103/PhysRevD.85.041301}{\emph{Phys.
  Rev.} {\bfseries D85} (2012) 041301}
  [\href{https://arxiv.org/abs/1106.3999}{{\ttfamily 1106.3999}}].

\bibitem{Bertacca:2015mca}
D.~Bertacca, N.~Bartolo, M.~Bruni, K.~Koyama, R.~Maartens, S.~Matarrese et~al.,
  \emph{{Galaxy bias and gauges at second order in General Relativity}},
  \href{https://doi.org/10.1088/0264-9381/32/17/175019}{\emph{Class. Quant.
  Grav.} {\bfseries 32} (2015) 175019}
  [\href{https://arxiv.org/abs/1501.03163}{{\ttfamily 1501.03163}}].

\bibitem{Umeh:2019qyd}
O.~Umeh, K.~Koyama, R.~Maartens, F.~Schmidt and C.~Clarkson, \emph{{General
  relativistic effects in the galaxy bias at second order}},
  \href{https://doi.org/10.1088/1475-7516/2019/05/020}{\emph{JCAP} {\bfseries
  1905} (2019) 020} [\href{https://arxiv.org/abs/1901.07460}{{\ttfamily
  1901.07460}}].

\bibitem{Umeh:2019jqg}
O.~Umeh and K.~Koyama, \emph{{The galaxy bias at second order in general
  relativity with Non-Gaussian initial conditions}},
  \href{https://doi.org/10.1088/1475-7516/2019/12/048}{\emph{JCAP} {\bfseries
  1912} (2019) 048} [\href{https://arxiv.org/abs/1907.08094}{{\ttfamily
  1907.08094}}].

\bibitem{Calles:2019prs}
J.~Calles, L.~Castiblanco, J.~Noreña and C.~Stahl, \emph{{From matter to
  galaxies: General relativistic bias for the one-loop bispectrum}},
  \href{https://arxiv.org/abs/1912.13034}{{\ttfamily 1912.13034}}.

\bibitem{Adamek:2016zes}
J.~Adamek, D.~Daverio, R.~Durrer and M.~Kunz, \emph{{gevolution: a cosmological
  N-body code based on General Relativity}},
  \href{https://doi.org/10.1088/1475-7516/2016/07/053}{\emph{JCAP} {\bfseries
  1607} (2016) 053} [\href{https://arxiv.org/abs/1604.06065}{{\ttfamily
  1604.06065}}].

\bibitem{Barrera-Hinojosa:2019mzo}
C.~Barrera-Hinojosa and B.~Li, \emph{{GRAMSES: a new route to general
  relativistic $N$-body simulations in cosmology - I. Methodology and code
  description}},  \href{https://arxiv.org/abs/1905.08890}{{\ttfamily
  1905.08890}}.

\bibitem{Bentivegna:2015flc}
E.~Bentivegna and M.~Bruni, \emph{{Effects of nonlinear inhomogeneity on the
  cosmic expansion with numerical relativity}},
  \href{https://doi.org/10.1103/PhysRevLett.116.251302}{\emph{Phys. Rev. Lett.}
  {\bfseries 116} (2016) 251302}
  [\href{https://arxiv.org/abs/1511.05124}{{\ttfamily 1511.05124}}].

\bibitem{jeong1}
D.~Jeong, F.~Schmidt and C.~M. Hirata, \emph{Large-scale clustering of galaxies
  in general relativity},
  \href{https://doi.org/10.1103/PhysRevD.85.023504}{\emph{Phys. Rev. D}
  {\bfseries 85} (2012) 023504}.

\bibitem{durrer1}
C.~Bonvin and R.~Durrer, \emph{What galaxy surveys really measure},
  \href{https://doi.org/10.1103/PhysRevD.84.063505}{\emph{Phys. Rev. D}
  {\bfseries 84} (2011) 063505}.

\bibitem{antony}
A.~Challinor and A.~Lewis, \emph{Linear power spectrum of observed source
  number counts}, \href{https://doi.org/10.1103/PhysRevD.84.043516}{\emph{Phys.
  Rev. D} {\bfseries 84} (2011) 043516}.

\bibitem{chen}
S.~Chen and D.~J. Schwarz, \emph{{Fluctuations of differential number counts of
  radio continuum sources}},
  \href{https://doi.org/10.1103/PhysRevD.91.043507}{\emph{Phys. Rev.}
  {\bfseries D91} (2015) 043507}
  [\href{https://arxiv.org/abs/1407.4682}{{\ttfamily 1407.4682}}].

\bibitem{zalda}
J.~{Yoo}, A.~L. {Fitzpatrick} and M.~{Zaldarriaga}, \emph{{New perspective on
  galaxy clustering as a cosmological probe: General relativistic effects}},
  \href{https://doi.org/10.1103/PhysRevD.80.083514}{\emph{Phys. Rev. D}
  {\bfseries 80} (2009) 083514}
  [\href{https://arxiv.org/abs/0907.0707}{{\ttfamily 0907.0707}}].

\bibitem{cc1}
D.~Bertacca, R.~Maartens and C.~Clarkson, \emph{{Observed galaxy number counts
  on the lightcone up to second order: I. Main result}}, {\emph{Journal of
  Cosmology and Astroparticle Physics} {\bfseries 2014} (2014) 037}.

\bibitem{cc2}
D.~Bertacca, R.~Maartens and C.~Clarkson, \emph{{Observed galaxy number counts
  on the lightcone up to second order: II. Derivation}}, {\emph{Journal of
  Cosmology and Astroparticle Physics} {\bfseries 2014} (2014) 013}.

\bibitem{durrer2}
E.~Di~Dio, R.~Durrer, G.~Marozzi and F.~Montanari, \emph{{Galaxy number counts
  to second order and their bispectrum}},
  \href{https://doi.org/10.1088/1475-7516/2014/12/017,
  10.1088/1475-7516/2015/06/E01}{\emph{JCAP} {\bfseries 1412} (2014) 017}
  [\href{https://arxiv.org/abs/1407.0376}{{\ttfamily 1407.0376}}].

\bibitem{yoo1}
J.~Yoo, \emph{General relativistic description of the observed galaxy power
  spectrum: Do we understand what we measure?},
  \href{https://doi.org/10.1103/PhysRevD.82.083508}{\emph{Phys. Rev. D}
  {\bfseries 82} (2010) 083508}.

\bibitem{yoo2}
J.~Yoo and M.~Zaldarriaga, \emph{{Beyond the Linear-Order Relativistic Effect
  in Galaxy Clustering: Second-Order Gauge-Invariant Formalism}},
  \href{https://doi.org/10.1103/PhysRevD.90.023513}{\emph{Phys. Rev.}
  {\bfseries D90} (2014) 023513}
  [\href{https://arxiv.org/abs/1406.4140}{{\ttfamily 1406.4140}}].

\bibitem{yoo3}
J.~Yoo and F.~Scaccabarozzi, \emph{{Unified Treatment of the Luminosity
  Distance in Cosmology}},
  \href{https://doi.org/10.1088/1475-7516/2016/09/046}{\emph{JCAP} {\bfseries
  1609} (2016) 046} [\href{https://arxiv.org/abs/1606.08453}{{\ttfamily
  1606.08453}}].

\bibitem{nielsen}
J.~T. Nielsen and R.~Durrer, \emph{{Higher order relativistic galaxy number
  counts: dominating terms}},
  \href{https://doi.org/10.1088/1475-7516/2017/03/010}{\emph{JCAP} {\bfseries
  1703} (2017) 010} [\href{https://arxiv.org/abs/1606.02113}{{\ttfamily
  1606.02113}}].

\bibitem{yoo5}
G.~Fanizza, J.~Yoo and S.~G. Biern, \emph{{Non-linear general relativistic
  effects in the observed redshift}},
  \href{https://arxiv.org/abs/1805.05959}{{\ttfamily 1805.05959}}.

\bibitem{Fuentes:2019nel}
J.~L. Fuentes, J.~C. Hidalgo and K.~A. Malik, \emph{{Galaxy number counts at
  second order: an independent approach}},
  \href{https://arxiv.org/abs/1908.08400}{{\ttfamily 1908.08400}}.

\bibitem{Umeh:2016nuh}
O.~Umeh, S.~Jolicoeur, R.~Maartens and C.~Clarkson, \emph{{A general
  relativistic signature in the galaxy bispectrum: the local effects of
  observing on the lightcone}},
  \href{https://doi.org/10.1088/1475-7516/2017/03/034}{\emph{JCAP} {\bfseries
  1703} (2017) 034} [\href{https://arxiv.org/abs/1610.03351}{{\ttfamily
  1610.03351}}].

\bibitem{Jolicoeur:2017nyt}
S.~Jolicoeur, O.~Umeh, R.~Maartens and C.~Clarkson, \emph{{Imprints of local
  lightcone \ projection effects on the galaxy bispectrum. Part II}},
  \href{https://doi.org/10.1088/1475-7516/2017/09/040}{\emph{JCAP} {\bfseries
  1709} (2017) 040} [\href{https://arxiv.org/abs/1703.09630}{{\ttfamily
  1703.09630}}].

\bibitem{Clarkson:2018dwn}
C.~Clarkson, E.~M. de~Weerd, S.~Jolicoeur, R.~Maartens and O.~Umeh, \emph{{The
  dipole of the galaxy bispectrum}},
  \href{https://doi.org/10.1093/mnrasl/slz066}{\emph{Mon. Not. Roy. Astron.
  Soc.} {\bfseries 486} (2019) L101}
  [\href{https://arxiv.org/abs/1812.09512}{{\ttfamily 1812.09512}}].

\bibitem{Mitsou:2019ocs}
E.~Mitsou, J.~Yoo, R.~Durrer, F.~Scaccabarozzi and V.~Tansella, \emph{{Angular
  $N$-point spectra and cosmic variance on the light-cone}},
  \href{https://arxiv.org/abs/1905.01293}{{\ttfamily 1905.01293}}.

\bibitem{Maartens:2019yhx}
R.~Maartens, S.~Jolicoeur, O.~Umeh, C.~Clarkson, S.~Camera and E.~M. De~Weerd,
  \emph{{Detecting the relativistic galaxy bispectrum}},
  \href{https://arxiv.org/abs/1911.02398}{{\ttfamily 1911.02398}}.

\bibitem{Bertacca:2017dzm}
D.~Bertacca, A.~Raccanelli, N.~Bartolo, M.~Liguori, S.~Matarrese and L.~Verde,
  \emph{{Relativistic wide-angle galaxy bispectrum on the light-cone}},
  \href{https://doi.org/10.1103/PhysRevD.97.023531}{\emph{Phys. Rev.}
  {\bfseries D97} (2018) 023531}
  [\href{https://arxiv.org/abs/1705.09306}{{\ttfamily 1705.09306}}].

\bibitem{Bartolo:2005xa}
N.~Bartolo, S.~Matarrese and A.~Riotto, \emph{{Signatures of primordial
  non-Gaussianity in the large-scale structure of the Universe}},
  \href{https://doi.org/10.1088/1475-7516/2005/10/010}{\emph{JCAP} {\bfseries
  0510} (2005) 010} [\href{https://arxiv.org/abs/astro-ph/0501614}{{\ttfamily
  astro-ph/0501614}}].

\bibitem{dePutter:2015vga}
R.~de~Putter, O.~Doré and D.~Green, \emph{{Is There Scale-Dependent Bias in
  Single-Field Inflation?}},
  \href{https://doi.org/10.1088/1475-7516/2015/10/024}{\emph{JCAP} {\bfseries
  1510} (2015) 024} [\href{https://arxiv.org/abs/1504.05935}{{\ttfamily
  1504.05935}}].

\bibitem{Bartolo:2015qva}
N.~Bartolo, D.~Bertacca, M.~Bruni, K.~Koyama, R.~Maartens, S.~Matarrese et~al.,
  \emph{{A relativistic signature in large-scale structure}},
  \href{https://doi.org/10.1016/j.dark.2016.04.002}{\emph{Phys. Dark Univ.}
  {\bfseries 13} (2016) 30} [\href{https://arxiv.org/abs/1506.00915}{{\ttfamily
  1506.00915}}].

\bibitem{Desjacques:2010jw}
V.~Desjacques and U.~Seljak, \emph{{Primordial non-Gaussianity from the large
  scale structure}},
  \href{https://doi.org/10.1088/0264-9381/27/12/124011}{\emph{Class. Quant.
  Grav.} {\bfseries 27} (2010) 124011}
  [\href{https://arxiv.org/abs/1003.5020}{{\ttfamily 1003.5020}}].

\bibitem{Bartolo:2004if}
N.~Bartolo, E.~Komatsu, S.~Matarrese and A.~Riotto, \emph{{Non-Gaussianity from
  inflation: Theory and observations}},
  \href{https://doi.org/10.1016/j.physrep.2004.08.022}{\emph{Phys. Rept.}
  {\bfseries 402} (2004) 103}
  [\href{https://arxiv.org/abs/astro-ph/0406398}{{\ttfamily
  astro-ph/0406398}}].

\bibitem{Giannantonio:2011ya}
T.~Giannantonio, C.~Porciani, J.~Carron, A.~Amara and A.~Pillepich,
  \emph{{Constraining primordial non-Gaussianity with future galaxy surveys}},
  \href{https://doi.org/10.1111/j.1365-2966.2012.20604.x}{\emph{Mon. Not. Roy.
  Astron. Soc.} {\bfseries 422} (2012) 2854}
  [\href{https://arxiv.org/abs/1109.0958}{{\ttfamily 1109.0958}}].

\bibitem{Taruya:2008pg}
A.~Taruya, K.~Koyama and T.~Matsubara, \emph{{Signature of Primordial
  Non-Gaussianity on Matter Power Spectrum}},
  \href{https://doi.org/10.1103/PhysRevD.78.123534}{\emph{Phys. Rev.}
  {\bfseries D78} (2008) 123534}
  [\href{https://arxiv.org/abs/0808.4085}{{\ttfamily 0808.4085}}].

\bibitem{Desjacques:2008vf}
V.~Desjacques, U.~Seljak and I.~Iliev, \emph{{Scale-dependent bias induced by
  local non-Gaussianity: A comparison to N-body simulations}},
  \href{https://doi.org/10.1111/j.1365-2966.2009.14721.x}{\emph{Mon. Not. Roy.
  Astron. Soc.} {\bfseries 396} (2009) 85}
  [\href{https://arxiv.org/abs/0811.2748}{{\ttfamily 0811.2748}}].

\bibitem{Tellarini:2016sgp}
M.~Tellarini, A.~J. Ross, G.~Tasinato and D.~Wands, \emph{{Galaxy bispectrum,
  primordial non-Gaussianity and redshift space distortions}},
  \href{https://doi.org/10.1088/1475-7516/2016/06/014}{\emph{JCAP} {\bfseries
  1606} (2016) 014} [\href{https://arxiv.org/abs/1603.06814}{{\ttfamily
  1603.06814}}].

\bibitem{DiDio:2016gpd}
E.~Di~Dio, H.~Perrier, R.~Durrer, G.~Marozzi, A.~Moradinezhad~Dizgah,
  J.~Noreña et~al., \emph{{Non-Gaussianities due to Relativistic Corrections
  to the Observed Galaxy Bispectrum}},
  \href{https://doi.org/10.1088/1475-7516/2017/03/006}{\emph{JCAP} {\bfseries
  1703} (2017) 006} [\href{https://arxiv.org/abs/1611.03720}{{\ttfamily
  1611.03720}}].

\bibitem{Koyama:2018ttg}
K.~Koyama, O.~Umeh, R.~Maartens and D.~Bertacca, \emph{{The observed galaxy
  bispectrum from single-field inflation in the squeezed limit}},
  \href{https://doi.org/10.1088/1475-7516/2018/07/050}{\emph{JCAP} {\bfseries
  1807} (2018) 050} [\href{https://arxiv.org/abs/1805.09189}{{\ttfamily
  1805.09189}}].

\bibitem{Akrami:2019izv}
{\scshape Planck} collaboration, \emph{{Planck 2018 results. IX. Constraints on
  primordial non-Gaussianity}},
  \href{https://arxiv.org/abs/1905.05697}{{\ttfamily 1905.05697}}.

\bibitem{Gressel:2017htk}
H.~A. Gressel and M.~Bruni, \emph{{f$_{NL}$-g$_{NL}$ mixing in the matter
  density field at higher orders}},
  \href{https://doi.org/10.1088/1475-7516/2018/06/016}{\emph{JCAP} {\bfseries
  1806} (2018) 016} [\href{https://arxiv.org/abs/1712.08687}{{\ttfamily
  1712.08687}}].

\bibitem{Bruni:2014xma}
M.~Bruni, J.~C. Hidalgo and D.~Wands, \emph{{Einstein's signature in
  cosmological large-scale structure}},
  \href{https://doi.org/10.1088/2041-8205/794/1/L11}{\emph{Astrophys. J.}
  {\bfseries 794} (2014) L11}
  [\href{https://arxiv.org/abs/1405.7006}{{\ttfamily 1405.7006}}].

\bibitem{Bruni:2013qta}
M.~Bruni, J.~C. Hidalgo, N.~Meures and D.~Wands, \emph{{Non-Gaussian Initial
  Conditions in $\Lambda$CDM: Newtonian, Relativistic, and Primordial
  Contributions}},
  \href{https://doi.org/10.1088/0004-637X/785/1/2}{\emph{Astrophys. J.}
  {\bfseries 785} (2014) 2} [\href{https://arxiv.org/abs/1307.1478}{{\ttfamily
  1307.1478}}].

\bibitem{Meures}
N.~Meures, \emph{{General Relativistic Effects on Cosmological Observations}},
  Ph.D. thesis, University of Portsmouth, 2012.

\bibitem{Malik:2008im}
K.~A. Malik and D.~Wands, \emph{{Cosmological perturbations}},
  \href{https://doi.org/10.1016/j.physrep.2009.03.001}{\emph{Phys. Rept.}
  {\bfseries 475} (2009) 1} [\href{https://arxiv.org/abs/0809.4944}{{\ttfamily
  0809.4944}}].

\bibitem{Carrilho:2015cma}
P.~Carrilho and K.~A. Malik, \emph{{Vector and tensor contributions to the
  curvature perturbation at second order}},
  \href{https://doi.org/10.1088/1475-7516/2016/02/021}{\emph{JCAP} {\bfseries
  1602} (2016) 021} [\href{https://arxiv.org/abs/1507.06922}{{\ttfamily
  1507.06922}}].

\bibitem{Salopek:1990jq}
D.~S. Salopek and J.~R. Bond, \emph{{Nonlinear evolution of long wavelength
  metric fluctuations in inflationary models}},
  \href{https://doi.org/10.1103/PhysRevD.42.3936}{\emph{Phys. Rev.} {\bfseries
  D42} (1990) 3936}.

\bibitem{Lyth:2004gb}
D.~H. Lyth, K.~A. Malik and M.~Sasaki, \emph{{A General proof of the
  conservation of the curvature perturbation}},
  \href{https://doi.org/10.1088/1475-7516/2005/05/004}{\emph{JCAP} {\bfseries
  0505} (2005) 004} [\href{https://arxiv.org/abs/astro-ph/0411220}{{\ttfamily
  astro-ph/0411220}}].

\bibitem{Lyth:2009zz}
D.~H. Lyth and A.~R. Liddle, \emph{{The primordial density perturbation:
  Cosmology, inflation and the origin of structure}}. Cambridge University
  Press, Cambridge U.K., (2009).

\bibitem{Lifshitz:1963ps}
E.~M. Lifshitz and I.~M. Khalatnikov, \emph{{Investigations in relativistic
  cosmology}}, \href{https://doi.org/10.1080/00018736300101283}{\emph{Adv.
  Phys.} {\bfseries 12} (1963) 185}.

\bibitem{Tomita:1975kj}
K.~Tomita, \emph{{Evolution of Irregularities in a Chaotic Early Universe}},
  \href{https://doi.org/10.1143/PTP.54.730}{\emph{Prog. Theor. Phys.}
  {\bfseries 54} (1975) 730}.

\bibitem{Rampf:2012pu}
C.~Rampf and G.~Rigopoulos, \emph{{Zel'dovich Approximation and General
  Relativity}}, \href{https://doi.org/10.1093/mnrasl/sls049}{\emph{Mon. Not.
  Roy. Astron. Soc.} {\bfseries 430} (2013) L54}
  [\href{https://arxiv.org/abs/1210.5446}{{\ttfamily 1210.5446}}].

\bibitem{Rampf:2013ewa}
C.~Rampf and G.~Rigopoulos, \emph{{Initial conditions for cold dark matter
  particles and General Relativity}},
  \href{https://doi.org/10.1103/PhysRevD.87.123525}{\emph{Phys. Rev.}
  {\bfseries D87} (2013) 123525}
  [\href{https://arxiv.org/abs/1305.0010}{{\ttfamily 1305.0010}}].

\bibitem{Deruelle:1994iz}
N.~Deruelle and D.~Langlois, \emph{{Long wavelength iteration of Einstein's
  equations near a space-time singularity}},
  \href{https://doi.org/10.1103/PhysRevD.52.2007}{\emph{Phys. Rev.} {\bfseries
  D52} (1995) 2007} [\href{https://arxiv.org/abs/gr-qc/9411040}{{\ttfamily
  gr-qc/9411040}}].

\bibitem{Wald:1984rg}
R.~M. Wald, \emph{{General Relativity}}. Chicago Univ. Pr., Chicago, USA, 1984,
  \href{https://doi.org/10.7208/chicago/9780226870373.001.0001}{10.7208/chicago/9780226870373.001.0001}.

\bibitem{Christopherson:2015ank}
A.~J. Christopherson, J.~C. Hidalgo, C.~Rampf and K.~A. Malik,
  \emph{{Second-order cosmological perturbation theory and initial conditions
  for $N$-body simulations}},
  \href{https://doi.org/10.1103/PhysRevD.93.043539}{\emph{Phys. Rev.}
  {\bfseries D93} (2016) 043539}
  [\href{https://arxiv.org/abs/1511.02220}{{\ttfamily 1511.02220}}].

\bibitem{Carlson:2009it}
J.~Carlson, M.~White and N.~Padmanabhan, \emph{{A critical look at cosmological
  perturbation theory techniques}},
  \href{https://doi.org/10.1103/PhysRevD.80.043531}{\emph{Phys. Rev.}
  {\bfseries D80} (2009) 043531}
  [\href{https://arxiv.org/abs/0905.0479}{{\ttfamily 0905.0479}}].

\bibitem{Djeong}
D.~Jeong, \emph{Cosmology with high ($z>1$) redshift galaxy surveys}, Ph.D.
  thesis, The University of Texas at Austin, 2010.

\bibitem{Scoccimarro:1996se}
R.~Scoccimarro and J.~Frieman, \emph{{Loop corrections in nonlinear
  cosmological perturbation theory 2. Two point statistics and
  selfsimilarity}}, \href{https://doi.org/10.1086/178177}{\emph{Astrophys. J.}
  {\bfseries 473} (1996) 620}
  [\href{https://arxiv.org/abs/astro-ph/9602070}{{\ttfamily
  astro-ph/9602070}}].

\bibitem{Lesgourgues:2011re}
J.~Lesgourgues, \emph{{The Cosmic Linear Anisotropy Solving System (CLASS) I:
  Overview}},  \href{https://arxiv.org/abs/1104.2932}{{\ttfamily 1104.2932}}.

\bibitem{Ade:2015xua}
{\scshape Planck} collaboration, \emph{{Planck 2015 results. XIII. Cosmological
  parameters}},
  \href{https://doi.org/10.1051/0004-6361/201525830}{\emph{Astron. Astrophys.}
  {\bfseries 594} (2016) A13}
  [\href{https://arxiv.org/abs/1502.01589}{{\ttfamily 1502.01589}}].

\bibitem{Blanchard:2019oqi}
{\scshape Euclid} collaboration, \emph{{Euclid preparation: VII. Forecast
  validation for Euclid cosmological probes}},
  \href{https://arxiv.org/abs/1910.09273}{{\ttfamily 1910.09273}}.

\bibitem{Chisari:2018vrw}
{\scshape LSST Dark Energy Science} collaboration, \emph{{Core Cosmology
  Library: Precision Cosmological Predictions for LSST}},
  \href{https://doi.org/10.3847/1538-4365/ab1658}{\emph{Astrophys. J. Suppl.}
  {\bfseries 242} (2019) 2} [\href{https://arxiv.org/abs/1812.05995}{{\ttfamily
  1812.05995}}].

\bibitem{Markovic:2019sva}
K.~Markovic, B.~Bose and A.~Pourtsidou, \emph{{Assessing non-linear models for
  galaxy clustering I: unbiased growth forecasts from multipole expansion}},
  \href{https://arxiv.org/abs/1904.11448}{{\ttfamily 1904.11448}}.

\bibitem{McEwen:2016fjn}
J.~E. McEwen, X.~Fang, C.~M. Hirata and J.~A. Blazek, \emph{{FAST-PT: a novel
  algorithm to calculate convolution integrals in cosmological perturbation
  theory}}, \href{https://doi.org/10.1088/1475-7516/2016/09/015}{\emph{JCAP}
  {\bfseries 1609} (2016) 015}
  [\href{https://arxiv.org/abs/1603.04826}{{\ttfamily 1603.04826}}].

\end{thebibliography}\endgroup
\end{document}